\documentclass[12pt]{article}
\newcommand{\sss}{\setcounter{equation}{0}}

\newtheorem{theorem}{THEOREM}[section]

\newtheorem{lemma}[theorem]{LEMMA}

\newtheorem{definition}[theorem]{DEFINITION}

\newfont{\BBFONT}{msbm10 scaled 1200}
\newcommand{\ERE}{\hbox{\BBFONT R}}
\newcommand{\ZETA}{\hbox{\BBFONT Z}}

\newcommand{\CE}{\hbox{\BBFONT C}}

\def\Bscr{{\mathcal B}}
\def\Dscr{{\mathcal D}}
\def\Hscr{{\cal H}}
\def\Fscr{{\cal F}}

\def\Scr{{\cal S}}
\def\Kscr{{\cal K}}
\def\Ascr{{\cal A}}
\def\Wscr{{\cal W}}
\def\Vscr{{\cal V}}
\def\Escr{{\cal E}}
\def\Wscr{{\mathcal W}}

\def\beq{\begin{equation}}
\def\ene{\end{equation}}
\def\bull{\begin{flushright} \vrule height 6pt width 6pt depth -.pt
\end{flushright}}

\begin{document}
\baselineskip=20 pt
\parskip 6 pt

\title{Inverse Scattering at a Fixed Quasi--Energy for Potentials Periodic in Time
\thanks{{\sc ams} 2000 classification 35P25, 35Q40, 81U40.
Research partially supported by project {\sc papiit--unam},  IN
101902.}}
\author{Ricardo Weder\thanks{Fellow Sistema Nacional de Investigadores.}\\
Instituto de Investigaciones en Matem\'aticas Aplicadas y en Sistemas.\\
Universidad Nacional Aut\'onoma de M\'exico, \\Apartado Postal 20-726, M\'exico D.F. 01000. M\'exico.\\
E-Mail: weder@servidor.unam.mx\\}
\date{}
\maketitle

\begin{center}
\begin{minipage}{6.0in}
\centerline{{\bf Abstract}}\bigskip We prove that the scattering
matrix at a fixed quasi--energy determines uniquely a
time--periodic potential that decays exponentially at infinity. We
consider potentials that for each fixed time belong to $L^{3/2}$
in space. The exponent $3/2$ is critical for the singularities of
the potential in space. For this singular class of potentials the
result is new even in the time--independent case, where it was
only known for bounded exponentially decreasing potentials.

\flushleft Short Title: Inverse Scattering at a Fixed
Quasi--Energy
\end{minipage}
\end{center}

\section{Introduction}\sss

We consider the scattering of a quantum-mechanical particle in $\ERE^3$ by its interaction with a short--range external potential that is periodic in time. The time--dependent Schr\"odinger equation is given by,

\beq i\frac{\partial}{\partial t} \varphi(t, x) = H (t) \varphi
(t, x), \varphi(t_0, x) = \varphi_0 (x), \label{1.1} \ene where
$H_0 = -\Delta$, and $H (t) = H_0 +V (t, x)$ are, respectively,
the unperturbed and the perturbed Hamiltonians.  Assuming that $V$
is real valued and that it satisfies appropriate conditions on its
regularity and on its decay as  $|x| \to \infty$, that we specify
below, and that it is periodic in time, with period  that we take
as $2\pi$, i.e., $V (t+2\pi,x) = V (t, x),$ the solution to
(\ref{1.1})  is given by a strongly continuous unitary group on
$L^2$,

\beq
\varphi (t) = U (t, t_0) \varphi_0, \quad \varphi_0 \in L^2,
\label{1.2}
\ene
where we denote by $\varphi(t)$ the function $\varphi (t,\cdot)$.

The wave operators with time lag $\tau \in \ERE$ are defined as
the following strong limits

\begin{equation}
W_\pm (\tau):= \hbox{s}-\lim_{t\rightarrow \pm\infty}U^\ast
(t+\tau,\tau) e^{-it H_0}. \label{1.3}
\end{equation}

We give below conditions assuring that the $W_\pm (\tau)$ exist and are complete, i.e., Range $ W_\pm(\tau)=$

$\Hscr_{ac}\left( U (\tau+2\pi,\tau)\right)$. Here $\Hscr_{ac}(U)$
denotes the subspace of absolute continuity of $U$. Then, the
scattering operators

\beq S(\tau) := W^{\ast}_+ (\tau) W_-(\tau), \quad \tau \in \ERE,
\label{1.4} \ene are unitary on $L^2$. Note that as $V$ is
periodic, $W_\pm (\tau+2 \pi) = W_\pm (\tau)$, and $ S(\tau+ 2\pi)
= S (\tau)$. Hence, it is  enough to consider $\tau \in [ 0,
2\pi)$. The construction of the scattering matrix associated to
$S(\tau)$ is a consequence of the application  to our problem of
the Howland--Floquet method \cite{8},\cite{41}, \cite{9} and of
the Kato--Kuroda scattering theory \cite{17}, \cite{20}. However,
to motivate physically the scattering matrix, and in particular,
to clarify  what is the meaning, in physical terms, of a
scattering experiment at a fixed quasi--energy, it is convenient
to briefly discuss how scattering experiments with $S (\tau)$ are
related to each other for different $\tau^\prime s$. Here we
follow \cite{23}. As is well known \cite{27}, the wave operators
satisfy the following intertwining relations,

\beq W_{\pm} (\tau)  = U (\tau, 0) W_\pm (0) e^{i\tau H_0}.
\label{1.5} \ene

Hence, the scattering operators satisfy,

\beq S (\tau) = e^{-i \tau H_0} S (0) e^{i \tau H_0}.
 \label{1.6}
\ene
The incoming asymptotic states $\varphi$ and $e^{i \tau
H_0}\varphi$ represent two identically prepared states, except for
a time lag $\tau$. Then, thinking in terms of the Heisenberg
representation of quantum mechanics, the operator $S(\tau)$
describes a scattering experiment corresponding to an incoming
asymptotic state prepared with a time lag $\tau$. Note that
particles that enter the interaction region at different times do
not interact with the same configuration of the potential. Hence,
to consider all possible scattering events we have to take into
account the whole family, $S(\tau), \tau \in[0, 2 \pi)$. A natural
way to do this is to let $S(\tau)$ act as a multiplication
operator in the enlarged space,

\beq \Hscr := L^2 (T, L^2, d t ) \label{1.7} \ene where $T$ is the
torus, $T : = \ERE /2\pi \ZETA$, with $\ZETA$ the integers, and
$dt$ the measure induced in $T$ by Lebesgue measure in $\ERE$.
That is to say, we consider the locally square--integrable
functions on $\ERE$ with values in $L^2$ that are periodic with
period $2\pi$,  with scalar product

\beq (\varphi, \psi)_\Hscr := \int_0^{2\pi}\, dt  \int_{\ERE^3}
\, dx \  \varphi (t, x) \overline{\psi (t, x)}. \label{1.7b} \ene

Hence, let us define the enlarged scattering operator, $\Scr$,

\beq (\Scr \varphi) (t, x) :=  S(t) \varphi (t, x), \, \varphi \in
\Hscr. \label{1.8} \ene $\Scr$ is unitary on $\Hscr$. Let us
denote,
\beq F_0 := - i\frac{\partial}{\partial t} + H_0.
\label{1.9} \ene

Taking the derivative with respect to $t$ in (\ref{1.8}) and using
(\ref{1.6}) see that formally,

\beq F_0 \, \Scr = \Scr \, F_0. \label{1.10} \ene Equation
(\ref{1.1}) can be considered as an approximation  to the
interaction of a quantum particle with an external quantum field.
In this approximation $-i \frac{\partial}{\partial t}$ is the
energy operator  for the external quanta (note that the spectrum
of $-i \frac{\partial}{\partial t}$ is $\ZETA$) and  $\Hscr$ is
the state space for the quanta and the particle. Furthermore,
$F_0$ is the total free energy operator, and $\Scr$ is the
scattering operator for the quanta and the particle. $F_0$ is
usually called the free quasi--Hamiltonian or the free Floquet
Hamiltonian. The commutation relation (\ref{1.10}) tells us that
the free quasi--energy is conserved in the scattering experiment.
In other words, in the scattering experiment the particle can gain
or loose energy only by absorbing or emitting quanta of the
external field. It is quite remarkable that on spite of the fact
that we consider the external field as a classical time--dependent
potential, $V$, the emitted and absorbed energy is quantized
\cite{27}. These considerations make it natural to define the
scattering matrix for $\Scr$ as the operator that is obtained by
diagonalizing $\Scr$ in a spectral representation of $F_0$ that
expresses the quanta and particle content of $F_0$ in a natural
way. This spectral representation is constructed as follows. For
any $n \in \ZETA$ denote $O_n = (n,\infty )$ and

\beq \hat\Hscr := \oplus_{n=-\infty}^{\infty} L^2 (O_n, L^2
\left(S^2_1\right), d \lambda ) . \label{1.11} \ene Then, by
taking Fourier series in $t$ and Fourier transform in $x$ we
obtain an unitary operator, $\Fscr_0$, from $\Hscr$ onto
$\hat\Hscr$, such that

\beq \Fscr_0  F_0 \Fscr_0^{-1}= \lambda, \label{1.12} \ene is the
operator of multiplication by the quasi--energy  $\lambda$ on
$\hat\Hscr$. Moreover, for $\lambda \in \ERE \backslash \ZETA$ we
denote,

\beq
\hat\Hscr (\lambda) := \oplus_{ m=-\infty}^{n} L^2 \left(
S^2_1\right ), \label{1.13} \ene
where $n$ is the only integer
such that, $n <\lambda < n + 1$. Note that,

\beq \hat\Hscr = \oplus \int^{+\infty}_{-\infty} \hat\Hscr
(\lambda) d \lambda. \label{1.14} \ene
We designate,

\beq
\hat\Scr := \Fscr_0 \Scr \Fscr_0^{-1}. \label{1.15} \ene

Hence, we prove that there is an unitary operator, $\hat \Scr
(\lambda)$, on $\hat \Hscr (\lambda) $ (see (\ref{4.80}),
(\ref{4.84}) and (\ref{4.85})) such that,

\beq \left( \hat \Scr \varphi\right) (\lambda) = \hat\Scr
(\lambda) \varphi (\lambda),
 \label{1.16}
 \ene
 for any $\varphi =
\varphi (\lambda) \in \oplus \int^{+\infty}_{-\infty} \hat \Hscr
(\lambda) d \lambda$. Furthermore, $\hat S (\lambda) = I + T
(\lambda)$, where $T (\lambda)$ is an integral operator in $\hat
\Hscr (\lambda)$. $\hat S (\lambda)$ is the scattering matrix and
the Hilbert--Schmidt integral kernels of $T (\lambda)$ are the
scattering amplitudes, both at a fixed quasi--energy $\lambda$. The
fact that $\hat S (\lambda)$ is an operator on $\hat \Hscr
(\lambda)$ exhibits the multi--channel nature of our scattering
process, where quanta of the external field are emitted or
absorbed by the particle.

The potentials $V(t, \cdot)\in L^{3/2}$ that  we consider are so
singular that the Hamiltonian $H(t)$ can not be defined as an
operator sum and we have to use quadratic form methods. Note,
however, that defining the Hamiltonian by quadratic form methods
is quite natural from the physical point of view, as what is
measured experimentally are the transition probabilities $\left(
H(t)\phi, \psi\right)$. Once we realize -in mathematical as well
as in physical grounds- that the Hamiltonian has to be defined by
quadratic form methods it is natural to assume that the potential
factorizes as $V= V_1\, V_2$, and to give our conditions on $V_j,
j=1,2$. This is also convenient since the singularities of the
potential make it necessary  to use the factorization method to
solve the direct scattering problem.

To motivate our conditions it is instructive to first consider the
case of time-independent potentials. By Sobolev's imbedding
theorem $W_{1} \subset L^6$. Moreover, multiplication by $V_j \in
L^3, j=1,2$, is a bounded operator from $L^6$ into $L^2$. As
$W_{1}$ is the quadratic form domain of the Laplacian, $V:= V_1
V_2$ is infinitesimally quadratic form bounded with respect to
$H_0$. This makes it possible to define the Hamiltonian $H_0+V$ by
quadratic form methods. If $V\in L^{3/2}$ we can take, $V_1:=
|V|^{1/2}$, and $V_2:= |V|^{1/2} \hbox{sign} V$. Each of the
inclusions above is sharp, and this is the reason why $3/2$ is the
critical exponent for the singularities of the potential. In the
time-periodic case we give conditions on $V$ that allow us to
adapt these estimates in a natural way. We assume that $V$
factorizes as follows,

\beq V(t,x)= V_1(t,x)\, V_2(t,x), \label{1.16b} \ene where \beq
V_1(t,x) =\sum\limits^{+\infty}_{m= - \infty} e^{imt}V_{1,m}(x),\,
V_2(t,x)=\left(\sum\limits^{+\infty}_{m= - \infty}
e^{imt}V_{2,m}(x)\right) V_3(t,x), \label{1.16c} \ene with
 \beq
\sum\limits^{+\infty}_{m=-\infty}
 \|V_{j,m}\|_{L^3} < \infty,
j=1,2, \label{1.16d} \ene
 $V_3(t,x)\in L^{\infty}\left(\ERE^4\right)$, and
$V_3(t+2\pi,x)= V_3(t,x)$. One possible choice is $V_1(t,x):=
|V(t,x)|^{1/2}, V_2(t,x):= |V(t,x)|^{1/2} \hbox{sign}V(t,x)$. As
mentioned above, in the time independent case these conditions are
satisfied if $V \in L^{3/2}$. Note that (\ref{1.16d}) is a
condition on the regularity in time of $\sum\limits^{+\infty}_{m=
- \infty} e^{imt}V_{j,m}(x), j=1,2$. In fact, there is a trade off
between the singularities of the potential in space and its
regularity in time. We take advantage of this trade off by
assuming that the potential is a product of a bounded function,
that is only measurable in time, and of two factors that can have
singularities in space of type $L^{3}$, but that  are regular
enough in time. To solve the inverse problem we further assume in
Theorem 1.1 that the potential decays exponentially.

The perturbed quasi--Hamiltonian, $F$, is a self--adjoint
extension of $F_0 + V$. Our main result is the following theorem.

\begin{theorem}

Suppose that $V$ is real valued and that  it factorizes as,
$V(t,x) = V_1(t,x)\, V_2(t,x)$, where for some $\delta_0
> 0, V_1(t,x) = e^{-\delta_0
|x|}\sum\limits^{+\infty}_{m= - \infty} e^{imt}V_{1,m}(x),
V_2(t,x)=  e^{-\delta_0 |x|}\left(\sum\limits^{+\infty}_{m= -
\infty} e^{imt}V_{2,m}(x)\right) V_3(t,x)$, with \break
$\sum\limits^{+\infty}_{m=-\infty}
 \|V_{j,m}\|_{L^3} < \infty,
j=1,2, \,V_3(t,x)\in L^{\infty}(\ERE^4)$, and $V_3(t+2\pi,x)=
V_3(t,x), t \in \ERE$. Then, the scattering matrix, $\hat \Scr
(\lambda)$, known at any fixed $\lambda \in \ERE \backslash \ZETA$
that is not an eigenvalue of $F$, determines uniquely the
potential $V$.

\end{theorem}
As we show in Theorem 4.3 if  $V$ is small enough, $F$ has no
eigenvalues. For a result on the absence of eigenvalues when $V$
is repulsive see \cite{38}, \cite{39}. For the exponential decay
of quasi--stationary states see \cite{44}. For the uniqueness of
the inverse scattering problem for $N$--body systems with
time--dependent potentials when the high--energy limit of the
scattering operator is known see \cite{36}.

The paper is organized as follows.

In Section 2 we construct the spectral representation of $F_0$.
Furthermore, we state results on the limiting absorption principle
(LAP) for $F_0$ that are an immediate consequence of the
corresponding results for the Laplacian in $\ERE^3$.

In Section 3 we construct the unitary propagator  $U (t, t_0)$.
Here we follow the method of Yajima \cite{43}, \cite{45}.
Actually, our result is an extension of  Yajima's \cite{43},
\cite{45} to the critical singularity $L^{3/2}$. This is possible
by the use of the end--point Strichartz estimate \cite{18}.

In Section 4 we prove the LAP for $F$, and we establish the
existence and completeness of the wave operators. Here we extend
the previous results of Yajima \cite{41} and Howland \cite{9}, to
the critical singularity $L^{3/2}$. Furthermore, we construct the
scattering matrix. We use the Howland--Floquet method \cite{8},
\cite{41}, \cite{9} and the Kato--Kuroda scattering theory
\cite{17}, \cite{20}.

In Section 5 we prove Theorem 1.1 adapting to our case the proof
of uniqueness at a fixed energy for bounded exponentially decreasing
time--independent potentials given in \cite{32} by Uhlmann and
Vasy. Here the estimates and the generalized limiting absorption
principle for  Faddeev's Green operator that we obtained in
\cite{34} play an essential role. For other results in uniqueness
at a fixed energy for exponentially decreasing time--independent
potentials see \cite{24}, \cite{3} and \cite{10}. For uniqueness
at a fixed energy of compactly supported perturbations of a
short--range potential see \cite{35}. For perturbed stratified
media see \cite{11}, \cite{37} and \cite{7}.

Finally, we briefly describe the Howland--Floquet method \cite{8},
\cite{41}, \cite{9} that  we use. Let us define the following
strongly--continuous unitary groups in $\Hscr,$

\beq \left( Y_0 (\tau) \varphi\right) (t) := e^{-i \tau H_0}
\varphi (t-\tau), \label{1.18} \ene

\beq (Y (\tau) \varphi) (t) := U (t, t-\tau) \varphi (t-\tau),
\tau \in \ERE. \label{1.19} \ene The generators of $Y_0$ and $Y$
are, respectively, $F_0$ and $F$, i.e., $Y_0(\tau) = e^{-i\tau
F_0}$, and $Y(\tau) = e^{-i\tau F}$, or in a more precise way, a
self--adjoint realization of the free and perturbed
quasi--Hamiltonians. Then,

\beq
\left( e^{i\tau F} e^{-i \tau F_0} \varphi \right) (t) = U (t, 0) U (0, t+\tau) e^{-i (\tau+t) H_0} e^{i t H_0}\varphi (t).
\label{1.20}
\ene
Hence, defining

\beq \Wscr_\pm := \hbox{s}-\lim_{\tau \to \pm \infty} e^{i \tau F}
e^{-i \tau  F_0}, \label{1.21} \ene and using (\ref{1.5}), we
obtain that

\beq (\Wscr_\pm \varphi) (t) = W_\pm (t) \varphi (t), \label{1.22}
\ene and then (see (\ref{1.8}))

\beq \Scr = \Wscr_+^\ast \Wscr_-.
 \label{1.23}
 \ene
This means that we can study the scattering theory for (\ref{1.1})
by studying  the wave operators $\Wscr_\pm$. Since now time is
another coordinate, we can apply to this extended scattering
problem the stationary theory of Kato and Kuroda \cite{17},
\cite{20}.  Note that $\Wscr_\pm$ and $\Scr$ are, respectively, the wave and the scattering
operators for the
quanta and the particle.

For applications to quantum mechanics and to atomic physics of the
scattering problem discussed in this paper see \cite{9.b} and
\cite{23}.
\section{The Free Quasi--Hamiltonian} \sss

We define $F_0$ as the following self--adjoint operator in
$\Hscr,$

\beq (F_0 \varphi) (t, x) := \left( - i \frac{\partial}{\partial
t}+ H_0 \right) \varphi (t, x) , \label{2.1} \ene with domain,

\beq D (H_0) := \{\varphi \in \Hscr: (- i \frac{\partial}{\partial
t} + H_0) \varphi \in \Hscr\}, \label{2.2} \ene with the
derivatives in distribution sense. By $\Fscr_s$ we denote the
Fourier series,

\beq (\Fscr_s \varphi)_m := \frac{1}{\sqrt {2 \pi}} \int^{2\pi}_0
\varphi (t) e^{-i m t} dt, \label{2.3} \ene as an unitary operator
from $L^2 (T)$ onto $\ell^2$ and by $\Fscr_T$ the Fourier
transform,

\beq (\Fscr_T \varphi) (k) :=  \frac{1}{(2\pi)^{3/2}}
\int^{+\infty}_{-\infty} e^{-i k x} \varphi (x) d x, \label{2.4}
\ene
as an unitary operator on $L^2$. Then,

\beq \tilde\Fscr := \Fscr_s \times \Fscr_T \label{2.5} \ene is
unitary from $\Hscr$ onto

\beq \tilde\Hscr : = \ell^2 (L^2). \label{2.6} \ene Clearly, \beq
F_0 = \tilde\Fscr^{-1} ( m+ k^{2}) \tilde \Fscr,
\label{2.7} \ene
and

\beq D (F_0) = \big\{ \varphi : (m+k^{2}) (\tilde\Fscr \varphi)_m
(k) \in \tilde\Hscr\big\}. \label{2.8} \ene
Then, $F_0$ is
absolutely continuous and its spectrum is $\ERE$. Define the
following unitary operator from $\tilde \Hscr$ onto $\hat\Hscr$
(see (\ref{1.11})),

\beq (\Fscr_r \varphi)_m (\lambda, \nu) := \frac{1}{\sqrt{2}}
(\lambda - m )^{\frac{1}{4}} \varphi_m \left(\sqrt{\lambda-m} \,
\nu\right), \lambda \in (m, \infty), \nu \in S^2_1, \ene
 and
designate,

\beq \Fscr_0 := \Fscr_r \tilde\Fscr. \label{2.9} \ene Then,

\beq \hat F_0 : = \Fscr_0 F_0 \Fscr_0^{-1} = \lambda, \label{2.10}
\ene is the operator of multiplication by the quasi--energy
$\lambda$ on $\hat\Hscr$, i.e., $\Fscr_0$ gives us the spectral
representation that we need. Observe that for $\varphi$ with
compact support,

\beq (\Fscr_0 \varphi)_m (\lambda, \nu) = \int \overline{\phi_m
(t, x,\lambda, \nu)}\varphi (t,x) dt \, dx,
 \label{2.10b} \ene
where $\phi_m$ is the following generalized eigenfunction of
$F_0$,

\beq \phi_m (t, x, \lambda, \nu) := \frac{1}{\sqrt{ 2}}
\frac{(\lambda-m)^\frac{1}{4}}{(2 \pi)^2} e^{imt} e^{i
(\lambda-m)^{1/2} \nu \cdot x}.
 \label{2.10c} \ene

For $s \in \ERE$ let us denote by $L^2_s$ the weighted $L^2$ space,

\beq L^2_s : = \left\{ \varphi \in \mathcal{D}' : ( 1+ x^2)^{s/2}
\varphi (x) \in L^2 \right\}, \label{2.11}\ene with norm,

\beq
\|\varphi\|_{L^2_s} = \| (1+x^2)^{s/2}\varphi (x) \|_{L^2}.
\label{2.12}
\ene

For $\rho > 0$ let $T(\rho)$ be the bounded trace operator from
$L^2_s, s > 1 / 2$, into $L^2 (S^2_1)$ such that,

\beq (T(\rho)\varphi) (\nu) = \rho \, (\Fscr_T\varphi) (\rho \,
\nu) \ , \ \varphi \in C^\infty_0. \label{2.13} \ene $T(\rho)$ has
the following properties (see for example \cite{21}, pages 4.20
and 4.26).
\begin{enumerate}
\item [1)] \beq T (0) : = \lim\limits_{\rho\downarrow 0} T (\rho)
= 0, \label{2.14} \ene
\end{enumerate}
\noindent where the limit exists in the operator norm.
\begin{enumerate}
\item[2)] \beq \|T(\rho)\|_{\Bscr(L^2_s, L^2 (S^2_1))} \leq C_s,
\rho \geq 0,
 \label{2.14b}
 \ene
 \beq
 \|T(\rho) - T (\rho^\prime)\|_{\Bscr(L^2_s, L^2
(S^2_1))} \leq C |\rho-\rho^\prime|^{s- 1 / 2}, 1 / 2 < s < 3/2,
\rho,\rho^\prime \geq 0. \label{2.15} \ene
\end{enumerate}
Let us denote

\beq \Hscr_s : = L^2 (T, L^2_s), \ s \in \ERE .
\label{2.16} \ene
We define,

\beq \left(T_m (\lambda) \varphi\right)  (\nu) :=  \frac{1}{\sqrt
2} \frac{1}{(\lambda-m)^{1/4}}
 \left[ T ((\lambda - m) ^{1/2} ) (\Fscr_s \varphi)_m \right] (\nu),
\label{2.17}
\ene
for $\lambda > m$, and

$$(T_m (\lambda) \varphi) (\nu) = 0, \hbox{ \ for \ } \ \lambda \leq m.$$

Define $\hat\Hscr (\lambda)$ as in (\ref{1.13}) and

\beq D (\lambda) :=  \oplus_{ m= - \infty}^n T_m (\lambda), \quad
n < \lambda < n+1.
 \label{2.18} \ene
Then, for $\lambda \in \ERE\backslash \ZETA , D (\lambda) \in
\Bscr (\Hscr_s, \hat\Hscr (\lambda)), s > 1 / 2$, and \beq
\|D(\lambda)\|_{ \Bscr (\Hscr_s, \hat\Hscr (\lambda))} \leq C
(1+|\lambda-n|^{\frac{s-1}{2}}), n < \lambda < n +1. \label{2.19}
\ene Denote,

\beq \hat\Hscr (\infty) := \oplus_{ m =-\infty}^{\infty} L^2
\left( S^2_1\right ). \label{2.20} \ene Hence, $\hat \Hscr
(\lambda) \subset \hat \Hscr (\infty)$, with the natural imbedding
where we take $\varphi_m\equiv 0$ for $ m>\lambda$. $D (\cdot)$ is
a locally H\"older continuous function from $\ERE \backslash
\ZETA$ into $\hat \Hscr (\infty) $ with exponent $s-1 / 2$, if $1
/ 2 < s < 3 / 2$. Moreover, if $1 < s < 3 / 2$ it extends to a
H\"older continuous function defined also for $\lambda$ integer,
but the exponent at any integer $\lambda$ is $\frac{s-1}{2}$.

Let us denote by $E_0$  the spectral family of $F_0$. Then,

\beq \Fscr_0 E_0 (\Delta) \varphi = \chi_\Delta (\lambda) D
(\lambda) \varphi, \varphi \in \Hscr_s, s > 1 / 2, \label{2.20b}
\ene for any Borel set $\Delta$, and where $\chi_O$ denotes the
characteristic function of any set $O \subset \ERE$.

Let $P_m$ be the following orthogonal projection operator,

\beq P_m \varphi = \frac{e^{imt}}{2 \pi} \int^{2\pi}_0 e^{-imt}
\varphi(t, x) \, dt. \label{2.21} \ene We have that,

\beq \Hscr =\oplus_{ m=-\infty}^{\infty} \Hscr_m , \hbox{ where }
\Hscr_m : = P_m \Hscr.
 \label{2.22} \ene
 For $z \in \CE_\pm $
denote,

\beq R_0 (z) := (F_0 - z)^{-1}, \label{2.23} \ene and

\beq r_0 (z) := (H_0 - z)^{-1}. \label{2.24} \ene
 Clearly,
 \beq
R_0 (z) =\oplus_{ m=-\infty}^{\infty} r_0 (z - m) P_m.
\label{2.24b} \ene

Let $W_{\alpha},\alpha\geq 0$, be the Sobolev space,

\beq W_\alpha : = \{ \varphi \in  L^2 : (1+k^2)^{\alpha / 2}
(\Fscr_T\varphi) (k) \in L^2\},
\label{2.25} \ene
with norm,

\beq \|\varphi\|_{W_\alpha}: = \| (1+k^2)^{\alpha/2} (\Fscr_T
\varphi) (k) \|_{L^2}. \label{2.26} \ene For $\alpha < 0,
W_{\alpha}$ is the dual of $W_{-\alpha} $ (with the pairing given
by the $L^2$ scalar product). Moreover, define,

\beq
W_{\alpha, s} : = \left\{ \varphi \in L^2_s : (1 + x^2)^{s/2}\varphi (x)\in {W_\alpha}\right\},
\label{2.27}
\ene
with norm

\beq \|\varphi\|_{W_{\alpha, s}}:= \| (1+x^2)^{s/2} \varphi
(x)\|_{W_\alpha}. \label{2.28} \ene
Finally, we designate,

\beq
\Kscr_{\alpha, s} := L^2 (T, W_{\alpha, s}).
\label{2.29}
\ene

The following results on the limiting absorption principle (LAP)
for $F_0$ are an immediate consequence of (\ref{2.24b}) and of the
well known results on the LAP for $H_0$, \cite{1}, \cite{5},
\cite{12}, \cite{13}, and \cite{21}. The following limits,

\beq R_{0,\pm}(\lambda ):=  \lim\limits_{\varepsilon \downarrow 0}
R_0 (\lambda \pm i \varepsilon) \label{2.30} \ene exist in the
uniform operator topology on $\Bscr (\Hscr_s, \Kscr_{1, - s}), s >
1 / 2$, for $\lambda \in \ERE \backslash \ZETA$. The convergence
is uniform for $\lambda$ in compact sets of $\ERE \backslash
\ZETA$ and the functions,

\beq R_{0,\pm} (\lambda) := \cases{ R_0  (\lambda),    & $\Im
\lambda \not= 0 $, \cr\cr R_{0,\pm}(\lambda) & , $ \Im \lambda=
0$,} \label{2.31} \ene
defined for $\lambda \in \CE_\pm \cup \ERE
\backslash \ZETA$ with values in $\Bscr (\Hscr_ s, \Kscr_{1, -
s})$ are analytic for $\Im\lambda \not = 0$ and locally H\"older
continuous for $\Im \lambda = 0$, with exponent, $\gamma$,
satisfying $\gamma < s- 1 / 2, 1 / 2 < s < 3 / 2$. If $s > 1,
R_{0,\pm}(\lambda)$ extend to H\"older continuous functions on
$\overline{\CE_\pm}$ but the exponent of H\"older continuity at
$\lambda \in \ZETA$ satisfies $\gamma <  s-1, 1  < s < 3 / 2$.
Furthermore, $R_{0,\pm }(\lambda)$ are bounded operators on
$\Bscr(\Hscr_s, \Kscr_{\alpha, -s}), 0 \leq \alpha \leq 1, s> 1 /
2$, for $\lambda \in \ERE \backslash \ZETA$ and if $s>1$ also at
$\lambda = \ERE$. Moreover, for any $\delta > 0$ there is a
constant $C_\delta$ such that

\beq \|R_{0,\pm} (\lambda) \| _{\Bscr (\Hscr_s, \,
\Kscr_{\alpha,-s})}\leq \frac{C_\delta}{\left(1+ \inf_{ n}
|\lambda-n|\right)^{\alpha-1}}, 0\leq \alpha \leq 1,
 \label{2.32} \ene
for all $\lambda \in \ERE$ with $\inf_{n} \  |\lambda - n| \geq
\delta$, and where if $s>1$ we can take $\delta = 0$. Moreover,
the $R_{0,\pm}(\lambda)$ are compact operators from $\Hscr_s$ into
$\Kscr_{\alpha, -s}, 0 \leq \alpha < 1$, with $ \lambda$ and $s$
as above.

It follows from (\ref{2.20b}) and the Stone's theorem that,

\beq \frac{d}{d\lambda} E_0 (\lambda) = \frac{1}{2\pi i} \bigg[
R_0 (\lambda + i0) - R_0 (\lambda-i0) \bigg] = D^\ast (\lambda) D
(\lambda), \lambda \in \ERE\backslash \ZETA,
 \label{2.33} \ene
 as a
bounded operator on $\Bscr (\Hscr_s, \Hscr_{-s}), s> 1 / 2$,  and
if $s>1$ also at $\lambda \in \ZETA$.

\section{The Unitary Propagator}\sss

Let us define the following class of potentials.

\begin{definition}
For any interval $I \subset \ERE$ we denote by $\Vscr (I)$ the
class of potentials $V (t, x), t\in I, x \in \ERE^3$, such that

\beq
V (t,x) = V_1 (t,x) + V_2 (t,x)
\label{3.1}
\ene
where,

\beq V_1 (t, x) \in L^\infty (I, L^{3/2}), \label{3.2} \ene and

\beq V_2 (t,x) \in L^1 (I, L^\infty). \label{3.3} \ene
\end{definition}

\bull
Note that $\Vscr (I)$ is a Banach space with the norm,

\beq \|V\|_{\Vscr(I)} : =\inf \left\{ \|V_1\|_{L^{\infty}(I,
L^{3/2})}+ \|V_2\|_{L^{1}(I, L^{\infty})} : V= V_1 + V_2 \right\}.
\label{3.3b} \ene The operator $(H_0+1)^{-1/2}$ is an integral
operator with kernel $G(x-y)$, where $G$ is the Bessel potential
that satisfies \cite{28},

\beq |G(x)|\leq C_a |x|^{-2} e^{-a|x|},\, \hbox{for all}\,\, a>1.
\label{3.4} \ene Then, by the H\"older and the generalized Young
inequalities \cite{25}

\beq \| |V_1 (t)|^{1/2} (H_0+1)^{-1/2}\|_{\Bscr (L^2)} \leq C \|
|V_1 (t)|^{1/2}\|_{L^{3}}. \label{3.5}
\ene Consider $g_n \in
C^\infty_0 (\ERE^3)$ such that $g_n \to |V_1(t)|^{1/2}$ in the
norm of $L^{3}$. By the Rellich local compactness  theorem $g_n
(H_0+1)^{-1/2}$ is compact in $L^2$, and furthermore, by
(\ref{3.5})

\beq \| (|V_1(t)|^{1/2}- g_n) (H_0 +1)^{-1/2}\|_{\Bscr(L^2)} \leq
C\| |V_1(t)|^{1/2}-g_n\|_{L^{3}}, \label{3.6} \ene
and it follows
that $|V_1(t)|^{1/2} (H_0+1)^{-1/2}$ is compact. Then, the
quadratic form,

\beq h_t (\varphi, \psi) = (H_0 \varphi, \psi) + (V(t) \varphi,
\psi), \label{3.7} \ene with domain $W_1$ is closed and bounded
below. Let $H(t)$ be the associated self--adjoint operator
\cite{25} for a.e. $t \in I$.

Let us consider the integral equation associated to (\ref{1.1})

\beq \varphi(t) = U_0 (t-t_0) \varphi_0 - i \int^t_{t_0} U_0
(t-\tau) V (\tau) \varphi (\tau) d\tau, \label{3.8} \ene where,

\beq U_0 (t-t_0) := e^{-i(t-t_0)H_0}. \label{3.9} \ene
We
construct below the solutions to (\ref{1.1}) by solving
(\ref{3.8}).  The key issue for this purpose is the following
end--point Strichartz estimates. Let us denote,

\beq (G_{t_0} \varphi) (t):= -i \int^t_{t_0} U_0 (t-\tau) \varphi
(\tau) d \tau. \label{3.10} \ene
Let $I$ be any interval in
$\ERE$, and denote,

\beq L^{p,q} (I):= L^q (I, L^p), 1\leq p, q \leq \infty.
\label{3.11} \ene The function $\varphi\in L^{p,q}_{loc} (I)$ if
$\varphi \in L^{p,q}(I^{\prime})$ for $I^{\prime}$ any compact
subinterval of I. Then \cite{18},

\beq e^{-itH_0} \in \Bscr (L^2, L^{6,2}(I)), \label{3.12} \ene
and
for $t_0 \in I$

\beq G_{t_0} \in \Bscr (L^{6/5, 2} (I), L^{6,2} (I)) \cap  \Bscr
(L^{6/5, 2}(I), C_b(I, L^2))\cap \Bscr (L^{2,1} (I), L^{6,2} (I)).
\label{3.13} \ene
Moreover, trivially,

\beq e^{-itH_0}\in \Bscr (L^2, C_b (I, L^2)), \label{3.14} \ene

\beq G_{t_0} \in \Bscr (L^{2,1} (I), C_b (I, L^2)), \label{3.15}
\ene where $C_b (I, L^2)$ denotes the continuous and bounded
functions from $I$ into $L^2$. Furthermore, the bounds on
(\ref{3.12}) - (\ref{3.15}) can be taken uniform on $t_0$ and $I$.
Let us designate

\beq \Ascr (I) : = C_b (I, L^2) \cap L^{6,2}(I), \label{3.16} \ene
with norm

\beq \|\varphi\|_{\Ascr (I)}:= \max \left[ \| \varphi\|_{C_b (I,
L^2)}, \|\varphi\|_{L^{6,2}(I)}\right], \label{3.17} \ene and

\beq \Ascr^{\prime} (I):= L^{2, 1} (I) + L^{6/5, 2} (I),
\label{3.18} \ene with norm,

\beq \|\varphi\|_{ \Ascr^\prime (I)}:= \inf \left\{
\|\varphi_1\|_{L^{2,1}(I)}+ \| \varphi_2\|_{L^{6/5,2}(I)} :
\varphi= \varphi_1 + \varphi_2\right\}. \label{3.19} \ene

We prepare the following result.

\begin{lemma}
Given $\varepsilon > 0$ there is a $\delta>0$ such that for any interval $I^\prime \subset I$, with $|I^\prime|\leq \delta,$
\end{lemma}

\beq \|V\|_{\Vscr(I^\prime)}\leq \varepsilon. \label{3.20} \ene

{\it Proof:} Take $V_{1,m} \in L^\infty (I, L^{3/2}\cap L^\infty)$
such that

\beq \|V_1 - V_{1,m}\|_{L^\infty (I, L^{3/2})} \leq
\frac{\varepsilon}{2}, \label{3.21} \ene
and denote \beq V_m = V_{1,m} + V_2. \label{3.22} \ene
Then,
\beq
\|V- V_m\|_{ \Vscr (I)} \leq \frac{\varepsilon}{2}. \label{3.23}
\ene
Moreover,

\beq \|V_m \|_{\Vscr (I^\prime)} \leq \|V_m\|_{L^1 (I^\prime,
L^\infty)}\leq \frac{\varepsilon}{2}, \label{3.24} \ene if
$|I^\prime| \leq \delta$ for $\delta$ small enough.
\bull
By
H\"older's inequality,

\beq \|V\|_{\Bscr (\Ascr (I), \Ascr^\prime (I))} \leq
\|V\|_{\Vscr(I)}.
\label{3.25} \ene
Moreover, by (\ref{3.13}) and
(\ref{3.15}),

\beq G_{t_0}\in \Bscr (\Ascr^\prime(I), \Ascr (I)) \label{3.26}
\ene
with bound uniform on $t_0$ and on $I$.  Denote,

\beq Q_{t_0}\varphi := G_{t_0} V\varphi. \label{3.27} \ene
Hence,
by (\ref{3.25}) and (\ref{3.26})

\beq \|Q_{t_0}\|_{\Bscr (A(I))} \leq C \|V\|_{\Vscr(I)},
\label{3.28} \ene where the constant $C$ is independent of $t_0$.
In consequence, by Lemma 3.2 for any $t_0 \in I$ there is a
$\delta
> 0$ such that for $I^\prime= [t_0 - \delta/2, t_0 + \delta /2],\,
Q_{t_0} $ is a contraction on $\Ascr (I^\prime)$ and then
(\ref{3.8}) has an unique solution on $\Ascr (I^\prime)$ given by

\beq
\varphi(t) = (I- Q_{t_0})^{-1} \ T_{t_0} \varphi_0, \   t \in I^\prime,
\label{3.29}
\ene
where

\beq T_{t_0}\varphi_0 = U_0 (t-t_0) \varphi_0. \label{3.30} \ene

The following Theorem is now proven as in the proof of Theorem 1
of \cite{45} (see also \cite{43}) by extending the solution given
by (\ref{3.29}) to $t\in I$ in successive steps of length
$\delta$. As by Sobolev's theorem $L^{6/5}$ is continuously
imbedded in $W_{-2}$, we have that $V \in \Bscr (L^6, W_{-2})$,
and then $H_0 +V \in \Bscr (L^2 \cap L^6, W_{-2})$. Moreover, for
$\varphi \in D (H(t)) \cap L^6$,

\beq H (t) \varphi = H_0 \varphi + V (t) \varphi. \label{3.29b}
\ene We  also use the notation $H (t)$ for $H_0 + V(t)$ when
viewed as a bounded operator from $L^2\cap L^6$ into $W_{-2}$.

\begin{theorem}

Suppose that $V$ satisfies $V (t+2\pi, x) = V(t,x),\, t \in \ERE,
x \in \ERE^3$ and that $V \in \Vscr ([ 0, 2 \pi])$. Then, there
exists a unique propagator  $U (t, t_0), (t,t_0) \in \ERE^2$ with
the following properties.

\begin{enumerate}
\item [1)] $U (t, t_0)$ is unitary in $L^2$ with $U (t, t_1) U
(t_1, t_0) = U (t, t_0), t_0, t_1, t\in \ERE$.

\item [2)] $U (\cdot, \cdot) $ is a strongly--continuous function
from $\ERE^2$ into $\Bscr (L^2)$.

\item [3)]
$U (t+ 2\pi, t_0 + 2 \pi) = U (t, t_0), t, t_0 \in \ERE$.

\item [4)] For any $t_0 \in \ERE, \varphi \in L^2, U (\cdot , t_0)
\varphi \in L^{6,2}_{loc}(\ERE)$ and it satisfies the equation

\beq U (t, t_0) \varphi_0 = U_0 (t-t_0) \varphi_0 - i \int^t_{t_0}
U_0 (t-\tau) V (\tau) U(\tau, t_0) \varphi_0 \, d \tau.
\label{3.31} \ene

\item [5)] There is a constant $C$ such that for all $t_0 \in
\ERE, \varphi_0 \in L^2$ and all bounded intervals $I\subset
\ERE$,

\beq
\|U (\cdot, t_0) \varphi_0\|_{L^{6,2 }(I)}\leq C (1+|I|)^{1/2} \| \varphi_0\|_{L^2}.
\label{3.32}
\ene

\item [6)] For any $t_0 \in \ERE$ and $\varphi_0 \in L^2, U
(\cdot, t_0) \varphi_0$ is a $W_{-2}$-- valued,
absolutely-continuous function and it satisfies the equation
(\ref{1.1}),

\end{enumerate}

\beq
i \frac{\partial}{\partial t} U (t, t_0) \varphi_0 = H (t) U (t, t_0) \varphi_0.
\label{3.33}
\ene

\end{theorem}
\bull
 Theorem 3.3 extends the results of \cite{43}, \cite{45} to
the critical singularity $L^{3/2}$. We could consider the problem
of the regularity of the propagator as in \cite{43} and \cite{45}.
Also, as in Theorem 1 of \cite{45} we could study the case of $V
\in L^\infty (\ERE, L^{3/2}) + L^1 (\ERE, L^\infty)$. We do not go
in these directions here. For other results on the unitary
propagator for time--dependent potentials see \cite{14},
\cite{16}, \cite{25}, \cite{30}, and the references quoted in
these works.

\section{The Limiting Absorption Principle} \sss

In this section we always  assume that $V$ is real valued, and
that $ V(t,x) = V_1(t,x)\, V_2(t,x)$, where $V_1(t,x) =
(1+|x|)^{-(1+\epsilon)/2}\sum\limits^{+\infty}_{m= - \infty}
e^{imt}V_{1,m}(x), V_2(t,x)= (1+|x|)^{-(1+\epsilon)/2}
\left(\sum\limits^{+\infty}_{m= - \infty} e^{imt}V_{2,m}(x)\right)
V_3(t,x)$, with,

\beq
\sum\limits^{+\infty}_{m=-\infty}
 \|V_{j,m}\|_{L^3} < \infty,
j=1,2, \,V_3(t,x)\in L^{\infty}(\ERE^4), \hbox{and} V_3(t+2\pi,x)=
V_3(t,x), t \in \ERE, \label{4.1} \ene
for some $\epsilon >0$. We
could also add to $V$ a bounded short--range term, but for
simplicity, and since our aim is to solve the inverse problem for
exponentially decreasing potentials, we will not do so. Since $H
(t)$ is defined as a quadratic form, i.e., the perturbation $V(t)$
is only form bounded with respect to $H_0$, we find it convenient
to use the Kato--Kuroda theory \cite{17}, \cite{20} with the
factorization method.

Let us denote by $\chi_1(t,x)$ the characteristic function of the
support of $V_1(t,x)$. We define,

\beq q_1 (t,x) :=V_1(t,x)+ e^{-x^2}(1-\chi_1(t,x)),
 \label{4.2}
\ene

\beq q_2 (t, x) := V_2 (t, x) \chi_1(t,x). \label{4.3} \ene Let
$A$ and $B$ be the following maximal operators of multiplication
in $\Hscr$,

\beq A \varphi := q_1 (t, x) \,\varphi (t, x),
 \label{4.4} \ene

\beq B \varphi := q_2 (t, x) \, \varphi (t, x).
 \label{4.5} \ene
Estimating as in (\ref{3.5}) we prove that $A$ and $B$ are bounded from $\Kscr_{1,0}$ into
$\Hscr$. Observe that,

\beq V=B A = A B \in \Bscr \left(\Kscr_{1,0}, \Kscr_{-1,0}\right).
\label{4.5b} \ene We define $q_j (t, x), j=1,2,$ as above only to
simplify some of the proofs below. With this definition the range
of $A$ is dense in $\Hscr$. Let us denote by $R (z) : =
(F-z)^{-1}, \Im z \not = 0$, the resolvent of $F$. By functional
calculus for $\Im z
> 0$, \beq R (z) = i \int^\infty_0 e^{i z \tau} Y (\tau) d \tau.
\label{4.5c} \ene
 Then,

\beq (R (z) \varphi) (t) = i \int^t_{-\infty} e^{i z t } U (t,
\tau) e^{-i z \tau} \varphi (\tau) d\tau. \label{4.6} \ene
By
(\ref{4.6})

\beq \|(q_j R (z) \varphi) (t) \|_{L^2} \leq
\int^{2\pi}_{-\infty}\| q_j U (t, \tau) e^{-i z \tau}\varphi
(\tau)\|_{L^2} \, d \tau, 0 \leq t \leq 2 \pi,
\label{4.7} \ene
and by (\ref{3.32}) and H\"older's inequality,

\beq \| q_j R (z) \varphi \|_\Hscr \leq C \|q_j\|_{L^{3, \infty}}
\,\int^{2\pi}_{-\infty} e^{\Im z \tau} \|\varphi (\tau) \|_{L^2}
\, d \tau \leq \frac{C}{\Im z}\, \|q_j \|_{L^{3, \infty}} \,
\|\varphi\|_{\Hscr}. \label{4.8} \ene
Then, for $\Im z > 0, A \,
R_0 (z), B R_0 (z), A R (z)$, and $B R (z)$ are bounded in
$\Hscr$. We prove that they are also bounded in $\Hscr$ for $\Im z
< 0$ in a similar way. It follows from (\ref{3.31}), (\ref{4.6})
and a simple calculation (see the proof of Lemma 3.3 of \cite{41})
that,

\beq R (z) = R_0 (z) - (B R_0 (\bar z))^\ast A R (z), \Im z \not=
0. \label{4.9} \ene This implies, in particular, that $F$ is an
extension of $F_0 + B A = F_0 + AB.$

As is well known, $r_0 (z)$ is an integral operator with kernel
$h_0 (\sqrt{z}|x-y|)$, where

\beq h_0 ( \sqrt{z} |x| ) := \frac{1}{4\pi} \frac{ e^{i
\sqrt{z}|x|}}  {|x|}, \label{4.10} \ene where we take the branch
of the square root with $\Im \sqrt{z}\geq 0,  \ z \in \CE$.

\begin{lemma}
Let $f$ and $g$ satisfy,

\beq f (t, x) = \sum_{m=-\infty}^{\infty} e^{ i m t} f_m (x) ,
\sum^{+\infty}_{m = - \infty} \|f_m\|_{L^3} < \infty,
 \label{4.11}
\ene

\beq g (t,x) = \sum^{\infty}_{m=-\infty} e^{i m t} g_m (x),  \
\sum^{\infty}_{m= -\infty}\|g_m\|_{L^3} < \infty \label{4.12}.
\ene Then, \beq J_\pm (\lambda) := f R_{0, \pm} (\lambda) g,
\lambda \in \overline{\CE^\pm} \label{4.13} \ene are compact
operators on $\Hscr$. The $\Bscr
 (\Hscr)$-valued functions $J_\pm$ are analytic for $\lambda
\in \CE^\pm$ and continuous for $\lambda \in  \overline{\CE^\pm}$.
Furthermore,

\beq \lim_{|\Im  \lambda
| \to \infty } J_\pm (\lambda) = 0,
\label{4.13b} \ene where the limit holds in the operator norm in
$\Bscr (\Hscr)$.
\end{lemma}
\noindent{\it Proof:} Denote,

\beq \hat J_\pm (\lambda) : = \Fscr_s f \, R_{0,\pm} (\lambda)
\,g\, \Fscr^{-1}_s. \label{4.14} \ene It is enough to prove that
$\hat J_\pm (\lambda)$ have the properties stated on the Lemma as
an operator on $\tilde{\Hscr}:=\ell^2 (L^2)$. By (\ref{2.24b}) and
(\ref{4.10}),

\beq (\hat J_\pm (\lambda) \varphi)_n = \sum_{m=-\infty}^{\infty}
d_{\pm, n, m} (\lambda) \,\varphi_m , \label{4.15} \ene
where
$d_{\pm, n, m} (\lambda)$ is the integral operator on $L^2$ with
kernel

\beq d_{\pm, n, m} (\lambda, x, y) := \sum_{\ell=-\infty}^{\infty}
f_{n-\ell}{(x)} \, h_0 ( \sqrt{\lambda -\ell} \, |x-y|) g
(y)_{\ell-m}. \label{4.16} \ene By H\"older's and generalized
Young's inequalities \cite{25}, $d_{\pm, n, m} (\lambda)$ are
bounded in $L^2$ for $\lambda \in \overline{\CE}_\pm$ and,

\beq \|d_{\pm, n, m} (\lambda) \|_{\Bscr (L^2)} \leq C
\sum_{\ell=-\infty}^{\infty} \| f_{n-\ell} \|_{L^3} \|g_{\ell-m}
\|_{L^3}, \label{4.17} \ene
and since

\beq \sup_n \sum_{m=-\infty}^{\infty} \| d_{\pm, n, m} (\lambda)
\|_{\Bscr (L^2)} \leq C   \left( \sum_{\ell=-\infty}^{\infty} \|
f_\ell\|_{L^3} \right)  \sum_{r=-\infty}^{\infty} \| g_r\|_{L^3},
\label{4.18} \ene

\beq \sup_m \sum_{n=-\infty}^{\infty} \| d_{\pm, n, m} (\lambda)
\|_{\Bscr (L^2)}\leq C \left( \sum_{\ell=-\infty}^{\infty} \|
f_\ell\|_{L^3} \right) \sum_{r=-\infty}^{\infty} \| g_r\|_{L^3},
\label{4.19} \ene we have that,

\beq \| \hat J_\pm (\lambda) \|_{\Bscr (\tilde{\Hscr} )} \leq C
 \left( \sum_{\ell=-\infty}^{\infty} \|
f_\ell\|_{L^3} \right) \sum_{r=-\infty}^{\infty} \| g_r\|
_{L^3}.
 \label{4.20} \ene
 Let $h \in C^\infty_0 (\ERE^3)$ satisfy $\int h (x) d x =
1$, and denote $h_\ell (x) = \ell^3 h (\ell x)$. We designate,
$f_{m, r} (x): = f_m (x)$ if $ |x| \leq r, f_{m,r}(x) = 0$ if $|x|
\geq r,$ and

\beq f^{(\ell,r)}_m (x) = \int h_{\ell} (x-y) f_{m, r}(y) \, d y
\in C^\infty_0 (\ERE^3), \label{4.22} \ene

\beq g^{(\ell,r)}_m (x) = \int h_\ell (x-y) g_{m, r}(y) \, d y\in
C^\infty_0 (\ERE^3), \label{4.23} \ene with $g_{m, r}$ defined as
$f_{m,r}$. $f_{m}^{( \ell,r)}\to f_m, g_{m}^{ (\ell,r)} \to g_m$,
strongly in $L^3$, as $\ell, r \to \infty$. We define,

\beq f^{(\ell,r,p)} := \sum_{|m|\leq p} e^{imt}
f_m^{(\ell,r)};\,\, g^{(\ell,r,p )} := \sum_{|m|\leq p} e^{imt}
g^{(\ell,r)}_m. \label{4.24} \ene

As $f^{(l,r,p)}$ and $g^{(l,r,p)}$ are bounded and have compact
support in $x$ and as $R_{0,\pm }(\lambda)$ are compact operators
from $\Hscr_s$ into $\Hscr_{-s}, s > 1/2$, the operators
$f^{(l,r,p)}\, R_{0,\pm}(\lambda)\, g^{(l,r,p)}$ are compact in
$\Hscr$, for $ \lambda \in \overline{\CE_{\pm}}$, and then, we
have that,

\beq \hat J_\pm^{(\ell,r, p)} (\lambda) : = \Fscr_s f^{(\ell,r,p)}
R_{0,\pm} (\lambda) g^{(\ell,r,p)} \Fscr^{-1}_s \label{4.25} \ene
are compact in $\tilde{\Hscr}$ for  $ \lambda \in
\overline{\CE_{\pm}}$. By (\ref{4.20}),

\beq \lim_{\ell,r,p \to  \infty} \hat J^{(\ell,r,p)}_\pm (\lambda)
= \hat J_\pm (\lambda), \label{4.26} \ene in the uniform operator
topology on $\Bscr (\tilde{\Hscr)}$, and hence, $\hat J_\pm
(\lambda)$ are compact for $\lambda \in \overline{\CE_\pm}$.
Moreover, as $F_0$ is self--adjoint,

\beq \| \hat J^{(\ell,r,p)}_\pm (\lambda)\|_{\Bscr
(\tilde{\Hscr})} \leq \frac{C}{|\Im \lambda|} \|f^{(\ell,r,p)}
\|_{L^\infty \left(T \times \ERE^3\right)}\|
g^{(\ell,r,p)}\|_{L^\infty \left(T \times \ERE^3\right)} ,
\label{4.27} \ene and as the limit in (\ref{4.26}) is uniform for
$\lambda \in \overline{\CE_{\pm}}$, (\ref{4.13b}) holds.

\bull
Let us denote,
\beq Q_{0, \pm}(\lambda) := B R_{0,\pm}
(\lambda) A, \ \lambda \in \overline{\CE_\pm}. \label{4.32} \ene
Since the multiplication operators by $\chi_1$ and by $V_3$ are
bounded on $\Hscr,  Q_{0, \pm}(\lambda) $ have all the properties
stated in Lemma 4.1. Denote,

\beq G_{0,\pm} (\lambda) := I + Q_{0, \pm} (\lambda), \lambda \in
\overline{\CE_\pm}.
 \label{4.32b} \ene
 It follows from (\ref{4.9}) and a
simple calculation (see \cite{15}, \cite{20} and \cite{41}) that

\beq B R (z) = G_{0, \pm}(z)^{-1} B R_0 (z), \ z \in \CE_\pm,
\label{4.33} \ene

\beq R (z) = R_0 (z) - R_0 (z) A G_{0, \pm} (z)^{-1} B R_0 (z), z
\in \CE_\pm, \label{4.34} \ene

\beq R (z) A = R_0 (z) A G_{0, \pm}(z)^{-1} \ , \ z\in \CE_\pm.
\label{4.35} \ene In (\ref{4.34}), (\ref{4.35}), by $R_0 (z) A $
and $R (z) A$ we actually mean the closure of these operators that
are bounded in $\Hscr$.
 These formulae are first proven for $\Im z$ large enough using (\ref{4.13b}) and then, they are extended to $z \in \CE_\pm$ by the
 analyticity of the resolvents. In particular, it follows that $G_{0,\pm}(z)$ is invertible in
$\Bscr (\Hscr)$ for $z \in \CE_\pm$. Denote,

\beq Q_\pm (z) := B R (z) A, \ z \in \CE_\pm, \label{4.36} \ene
and

\beq G_\pm (z) := (I-Q_\pm (z)), z \in \CE_\pm. \label{4.37} \ene
Then,

\beq G_\pm (z) = G_{0, \pm} (z)^{-1},  \ z \in \CE_\pm.
\label{4.38} \ene To obtain the LAP for $F$ we need to prove that
$G_{0, \pm} (\lambda)$ are invertible for $\lambda \in \ERE
\backslash \ZETA$ if $\lambda$ is not an eigenvalue of $F$. This
is done extending the well known argument of \cite{1} where the
case of time--independent potentials was considered. This was
accomplished in \cite{5} and \cite{9}, but as they  considered the
case where the perturbation is relatively bounded, and the
factorization method is not needed, we give some details in the
Lemma below. We denote by $\sigma_p (F)$ the set of all
eigenvalues of $F$.

\begin{lemma}

Suppose that (\ref{4.1}) holds. Then, $Q_{0, \pm} (\lambda)$ are
invertible in $\Bscr(\Hscr)$ for $\lambda \in\ERE \backslash
\ZETA$ if and only if $\lambda$ is not an eigenvalue of $F$.

\end{lemma}
{\it Proof:} We first assume that $\lambda \in (\ERE \backslash
\ZETA) \cap \sigma_p (F)$, and that $F \varphi = \lambda \varphi$.
Taking the adjoint of (\ref{4.35}) we have that,

\beq (I + Q_{0, \mp}^\ast ( \bar{z})) A R (z) = A R_0 (z), z \in
\CE_\pm. \label{4.40} \ene Then (recall that $D (A)\supset D (F)
)$, \beq (I + Q^\ast_{0,\mp}( \bar{z})) A \varphi = A R_0 (z)
(F-z) \varphi. \label{4.41} \ene Take now $z = \lambda +i
\varepsilon$. Hence,

\beq \lim_{\varepsilon \downarrow 0} A R_0 ( \lambda +
i\varepsilon) (F-\lambda -i\varepsilon)\varphi =
-\lim_{\varepsilon \downarrow 0} A R_0 (\lambda + i \varepsilon) i
\varepsilon \varphi = 0, \label{4.42} \ene where the limit holds
in the strong topology of $\Kscr_{-1,(1+\varepsilon)/2 }$. Here I
use that $A \in \Bscr (\Hscr, \Kscr_{-1, (1+\varepsilon)/2})$ and
that $\varepsilon R_0 (\lambda + i \varepsilon) \to 0 $ as $
\varepsilon \to 0$, in the strong topology in $\Hscr$ because
$F_0$ has no eigenvalues. Then,

\beq (I + Q_{0, \pm}^\ast (\lambda)) A \varphi = 0, \label{4.43}
\ene and as $A \varphi \in \Hscr, \, (I + Q^\ast_{0, \pm}
(\lambda))$ are not invertible. Since $Q_{0, \pm} (\lambda) $ are
compact it follows that $(I + Q_{0, \pm}(\lambda))$ are not
invertible in $\Bscr (\Hscr)$.

Suppose now that $(I + Q_{0,\pm} (\lambda))$ are not invertible in
$\Bscr(\Hscr)$. Then, also $(I + Q_{0, \pm}^\ast (\lambda))$ are
not invertible and there are $w_\pm \in \Hscr$ such that

\beq w_\pm + A R_{0, \pm} (\lambda) B w_\pm = 0, w_\pm \not = 0.
\label{4.44} \ene
Taking the inner product of (\ref{4.44}) with $B
R_{0, \pm} (\lambda) B w_\pm$ we obtain that

\beq (w_\pm, B R_{0, \pm} (\lambda) B w_\pm ) + (A R_{0, \pm}
(\lambda) B w_\pm, B R_{0, \pm } (\lambda) B w_\pm )= 0.
\label{4.45} \ene Taking the imaginary part of (\ref{4.45}) and
using (\ref{2.33}) we have that

\beq D (\lambda) B w_\pm = 0. \label{4.46} \ene
Note that by
(\ref{2.33}) and Lemma 4.1, $ D (\lambda) B \in \Bscr (\Hscr, \hat
\Hscr (\lambda))$. Denote,

\beq \varphi_\pm : = - R_{0, \pm} (\lambda) B w_\pm. \label{4.47}
\ene $\varphi_\pm \not= 0$, because otherwise $w_\pm =A
\varphi_\pm= 0$. Designate, $\varphi_{\pm, m} : = (\Fscr_s
\varphi_\pm ) _m \, , \, (B w_\pm )_m : = (\Fscr_s B w_\pm)_m$.
Then, by (\ref{2.24b}),

\beq \varphi_{\pm, m} = - r_{0, \pm}(\lambda-m) (B w_{\pm})_m.
\label{4.48} \ene
The following statements are a slight extension
of the results of \cite{1} and \cite{5} that consider the case of
$\varphi \in W_{s,0}$. They are proven as in \cite{1}, \cite{5}.
\begin{enumerate}
\item[{\bf 1})] Let $c > 0$ and $s \in \ERE$. Then, for some
constant $C$, for all $\varphi \in W_{s,-1}$, \beq \left\|
\frac{\varphi (k)}{k^2 + \lambda^2}\right\|_{W_{s, 0}} \leq
\frac{C}{\lambda} \left\|\varphi\right\|_{W_{s, -1}}, \lambda
> c. \label{4.49} \ene
\item[{\bf 2})] Let $c > 0$ and $s > 1 / 2 $. Then, for some
constant $C$, for all $\varphi \in W_{s,-1}$ with $\varphi
(k)|_{|k| =\lambda} = 0$ in trace sense,
\end{enumerate}

\beq \left\| \frac{\varphi (k)}{k^2 - \lambda^2}\right\|_{W_{s-1,
0}}\leq C \left\|\varphi\right\|_{W_{s, - 1}}, \lambda > c.
\label{4.50} \ene
 For $\lambda - m < 0$ we obtain from
(\ref{4.48}) and (\ref{4.49}), with $s=\frac{1+\varepsilon}{2}$,

\beq \left\|\varphi_{\pm, m}\right\|_{L^2_{2s}}  \leq
\frac{C}{|\lambda - m|^{1/2}} \|(B w_\pm)_m \|_{W_{-1, 2 s}},
\label{4.51} \ene and when $\lambda - m
> 0 $  by (\ref{4.48}) and (\ref{4.50}),

\beq \|\varphi_{\pm , m}\|_{L^2_{2 s-1}} \leq C \| (Bw_\pm)_m
\|_{W_{-1, 2s}}. \label{4.52} \ene
By Lemma 4.1 and (\ref{4.44})
$w_\pm \in \Hscr_s,$ and,

\beq \|w_\pm\|_{\Hscr_s} \leq C \| w_\pm\|_{\Hscr_{-s}}. \label{4.53}
\ene
Equations (\ref{4.51}), (\ref{4.52}) and (\ref{4.53}) imply
that

\beq \|\varphi_\pm \|_{\Hscr_{2 s-1}} \leq C \|B w_\pm
\|_{\Kscr_{-1, 2s}}\leq C \|w_\pm\|_{\Hscr_s}\leq C \| w_\pm
\|_{\Hscr_{-s}}, \, 2s-1=\varepsilon
> 0. \label{4.54} \ene We now prove that the  $\varphi_\pm$ are
eigenvectors  of $F$ with eigenvalue $\lambda$. Taking the adjoint
of (\ref{4.35}) we have that

\beq A R (z) = (I + A R_0 (z) B)^{-1} A R_0 (z), z \in \CE_\pm.
\label{4.55} \ene
Taking the inverse of (\ref{4.55}) and
multiplying  the result by $A$ we obtain

\beq (F-z) = (F_0-z) A^{-1} (I + A R_0 (z) B) A, z \in \CE_\pm.
\label{4.56} \ene
Taking the limit $z \to \lambda$,

\beq F-\lambda = (F_0 - \lambda) A^{-1} (I + A R_{0, \pm}
(\lambda) B) A. \label{4.57} \ene
Then, since $\varphi_\pm \in D
(A)$ and $ A \varphi_\pm = w_\pm$, it follows from (\ref{4.44})
and (\ref {4.57}) that $\varphi_\pm \in D (F)$ and $F\varphi_\pm =
\lambda \varphi_\pm$.

\bull

The LAP for $F$ follows now from the LAP for $F_0$ (see Section 2) (\ref{4.34}) and Lemma 4.2.
We state the results in the following theorem.
\begin{theorem}
Suppose that (\ref{4.1}) holds. Then, the following limits

\beq R_\pm (\lambda ):= \lim_{\varepsilon \downarrow 0} R (\lambda
\pm i \varepsilon), \label{4.58} \ene exist in the uniform
operator topology on $\Bscr (\Hscr_s, \Hscr_{-s}), s =
\frac{1+\varepsilon}{2}$, for $\lambda \in \ERE \backslash \ZETA
\backslash \sigma_p (F)$, where $\sigma_p (F)$ denotes the set of
eigenvalues of $F$. Furthermore,

\beq R_\pm (\lambda) = R_{0, \pm} (\lambda) - R_{0, \pm} (\lambda)
A \, G_{0, \pm} (\lambda)^{-1} B R_{0, \pm}(\lambda).
\label{4.59}\ene The functions

\beq R_\pm (\lambda) := \cases{ R (\lambda), & $\lambda \in
\CE_\pm$, \cr R_\pm (\lambda ), & $\lambda \in \ERE \backslash
\ZETA \backslash \sigma_p (F) $,\cr} \label{4.60} \ene with values
in $\Bscr (\Hscr_s, \Hscr_{-s})$ are analytic for $\lambda \in
\CE_\pm$ and continuous for $\lambda \in \CE_\pm \cup \ERE
\backslash \ZETA \backslash \sigma_p (F)$. Furthermore, $F$ has no
singular-continuous spectrum and $\sigma_p (F) \backslash \ZETA$
consists of finite dimensional eigenvalues that  can only
accumulate at $\ZETA$. Moreover, if $\lambda \in \sigma_p (F),
\lambda + m \in \sigma_p (F)$ for all $m \in \ZETA$. Finally, if
either $\sum_{m=-\infty}^{\infty}  \| V_{j,m} \|_{L^3}$, for
$j=1$, or for $j=2$ or $\|V_3\|_{L^{\infty} \left(\ERE^4 \right)}$
is small enough, then, $\sigma_p (F)$ is empty.
\end{theorem}

\noindent{\it Proof:} The existence of the limits in (\ref{4.58})
and the properties of $R_\pm$ have already been proven. The fact
that $F$ has no singular--continuous spectrum is a consequence of
the LAP for $F$ \cite{26}. In the proof of Lemma 4.2 we
established a one--to--one correspondence between the kernel of
$F-\lambda$ and the kernel of $(I+ Q_{0, \pm} (\lambda))$ for
$\lambda \in \ERE \backslash \ZETA$. But as $Q_{0, \pm}(\lambda)$
are compact the kernel of the later are finite dimensional, and then
the non--integer eigenvalues of $F$ have finite multiplicity.
Suppose that $\lambda_j$ are infinite distinct points of $\sigma_p
(F) \backslash \ZETA$, and that $\lim\limits_{j \to
\infty}\lambda_j = \lambda_\infty, \lambda_\infty \not\in \ZETA$.
By Lemma 4.2 there are $w_j$ with $\|w_j \|_\Hscr = 1$
 such that

\beq w_j = - A R_{0, +} (\lambda_j) B w_j, j= 1,2, \cdots.
\label{4.61} \ene As $A R_{0, +} (\lambda_ j) B$ are compact
 we can assume (eventually passing to a subsequence) that $w_j \to
w_\infty$ strongly in $\Hscr$ with $\|w_\infty\|_\Hscr = 1$, and

\beq w_\infty = - A R_{0, +} (\lambda_\infty) B w_\infty.
\label{4.62} \ene
Denote by

\beq \varphi_j := - R_{0, +} (\lambda_j) Bw_j, \, j = 0, 1, 2,
\ldots, \infty, \label{4.63} \ene the corresponding sequence of
eigenvectors. Then $\varphi_j \not= 0$ and by (\ref{4.54})

\beq \|\varphi_j\|_\Hscr \leq C , j = 1, 2, \ldots, \infty.
\label{4.64} \ene
Since $w_j \to w_\infty$ strongly in $\Hscr$, by
(\ref{4.63}) $\varphi_j \to \varphi_\infty$ strongly in
$\Hscr_{-s}, s= \frac{1+\varepsilon}{2}$, to $\varphi_\infty$, and
then, by (\ref{4.64}) $\varphi_j \to \varphi_\infty$ weakly in
$\Hscr$. But as $\varphi_\infty$ and $\varphi_j$ are eigenvectors
corresponding to different eigenvalues of $F$ they are orthogonal,
and it follows that $\varphi_{\infty} = 0$, which is a
contradiction. The last statement of the theorem is immediate
since $e^{-imt} F e^{imt}= F + m$.
 Finally, by the proof of Lemma 4.1 if either $\sum\limits^{+\infty}_{m = - \infty}
\|V_{j,m}\|_{L^3}$ for $j=1$, or for $j=2$ or $\|V_3\|_{L^{\infty}\left(\ERE^4\right)}$ is small
enough $\|Q_{0,\pm} (\lambda) \| < 1$,
 and then, $I+ Q_{0,\pm}(\lambda)$ is invertible for all $\lambda \in \ERE$. Hence
(see (\ref{4.34}) and (\ref{4.60})), the LAP for $F$
 holds  for all   $\lambda$ in $\ERE$ and this implies that $F$ has pure absolutely-continuous
  spectrum \cite{26}. In particular,
 $F$ has no eigenvalues.

\bull
 Theorem 4.3 extends the results of \cite{9} and \cite{41} to
the critical singularity $L^{3/2}$. There are a number of results
in the LAP for time--dependent Hamiltonians. See for example,
\cite{22}, \cite{46}, where long--range potentials are considered.
In \cite{5} the LAP at $\lambda \in \ZETA$ is also studied.
References \cite{22}, \cite{46} and \cite{5} consider potentials
that are more regular than ours.

It follows from (\ref{2.33}) and Lemma 4.1 that $D (\lambda) A \in
B (\Hscr,\hat\Hscr (\lambda))$. Under the assumptions of Theorem
4.3 $G_\pm (\lambda)$ extend to continuous functions from $\CE_\pm
\cup
 \ERE \backslash\ZETA \backslash  \sigma_p (F)$ into  $\Bscr (\Hscr)
$ and

\beq G_\pm (\lambda)= G_{0, \pm} (\lambda)^{-1}, \lambda \in
\CE_\pm \cup \ERE \backslash\ZETA \backslash \sigma_p (F).
\label{4.65} \ene Denote,

\beq D_\pm(\lambda) : = D (\lambda) A G_\pm (\lambda), \lambda \in
\ERE \backslash \ZETA \backslash \sigma_p (F). \label{4.66} \ene
Let $E (\lambda)$ be the spectral family of $F$, and let
$\Hscr_{ac}(F)$ be the subspace of absolute continuity of $F$.
Then, for any $\varphi \in \Hscr_{ac} (F)$ of the form,

\beq \varphi = \sum^N_{j=1} E (I_j) A \varphi_ j, \bar I_j \subset
\ERE \backslash \ZETA \backslash \sigma_p (F), \label{4.67} \ene
$\bar I_j$ compact, $I_j \cap I_k = \phi, j \not= k$, define the
operators

\beq \Fscr_\pm \varphi := \sum^N_{j=1} \chi_{I_j} (\lambda) D_\pm
(\lambda) A \varphi_j. \label{4.68} \ene Hence, under (\ref{4.1})
the following facts are proven as in the proof of Theorem 3.11 of
\cite{20} (see also the proof of Lemma 7.1 of \cite{33}). The
$\Fscr_{\pm}$ defined by (\ref{4.68}) extend to unitary  operators
from $\Hscr_{ac} (F)$ onto $L^2 (\ERE, \hat\Hscr)$ and for any
$\varphi \in D (A)$ and any Borel set $I \subset \ERE \backslash
\ZETA \backslash \sigma_p (F)$,

\beq \Fscr_\pm E (I) A \varphi = \chi_I(\lambda) D_\pm (\lambda)
\varphi, \label{4.70} \ene
and if $P_{ac} (F)$ denotes the
orthogonal projector onto  $\Hscr_{ac} (F)$,

\beq F P_{ac}(F) = \Fscr^{-1}_\pm \lambda \, \Fscr_\pm.
 \label{4.71}
\ene

We now define the wave operators as

\beq \Wscr_\pm := \Fscr^{\ast}_\pm \Fscr_0. \label{4.72} \ene
The
$\Wscr_\pm$ are unitary from $\Hscr$ onto $\Hscr_{ac} (F)$.

Let us prove that the time--dependent formulae (\ref{1.21}) hold.
In fact, we shall prove a more general result known as the
invariance principle \cite{27}. Let $f (\lambda)$ be a
real--valued measurable function defined on $\ERE$ such that,

\beq \lim\limits_{\tau \to \infty} \int^\infty_0 \bigg|
\int_{\ERE} \varphi (\lambda) e^{-i\tau f(\lambda) -ix \lambda}
d\lambda \bigg|^2 d x = 0, \label{4.73} \ene for all $\varphi \in
L^2 (\ERE)$. Then,

\beq \Wscr_\pm = {\hbox{s}-\lim_{\tau \to \pm \infty}} e^{i \tau f
(F)} e^{- i \tau f (F_0)}. \label{4.74} \ene Note that $f
(\lambda) = \lambda$ satisfies (\ref{4.73}) and then, (\ref{1.21})
hold. We give some details of the proof in the $+$ case. By
unitarity it is enough to prove that

\beq \lim_{\tau \to \infty}(e^{-i \tau f (F)} E (I_1) A \varphi,
e^{-i \tau f (F_0)} E_0 (I_0) A \psi) = (E (I_1) A \varphi,
\Fscr^\ast_+ \Fscr_0 E_0 (I_0) A \psi), \label{4.75} \ene
for
$\varphi, \psi \in D ((1+|x|)^{(1+\varepsilon)/2}A)$ and $I_0,
I_1$ bounded intervals with $\bar I_0, \bar I_1 \subset \ERE
\backslash \ZETA \backslash \sigma_p (F)$.
But by (\ref{2.20b}),
(\ref{2.33}), (\ref{4.35}), (\ref{4.38}) and (\ref{4.70}),

$$
(e^{-i \tau f (F)} E (I_1) A \varphi, \ e^{-i \tau f (F_0)} E_0
(I_0) A \psi)
 = (E (I_1) A \varphi, \Fscr^\ast_+ \Fscr_0 E_0 (I_0) A \psi) +
$$
\beq
  \int_{I_1} e^{-i \tau f (\lambda)} \lim\limits_{\varepsilon \downarrow 0} h_{\varepsilon, \tau}  (\lambda) d \lambda,
 \label{4.76}
\ene
 where

\beq h_{\varepsilon, \tau} (\lambda) :=\frac{1}{2\pi i}  ( [ G_+
(\lambda + i \varepsilon) - G_- (\lambda - i \varepsilon)]
\varphi, A R_{0, +} (\lambda + i \varepsilon) e^{-i \tau f( F_0)}
E_0 (I_0) A \psi). \label{4.77} \ene
We prove that

\beq \lim_{\tau \to \infty} \int_{I_1} e^{-i \tau f (\lambda)}
\lim_{\varepsilon\downarrow 0} h_{\varepsilon, \tau}(\lambda) d
\lambda = 0, \label{4.78} \ene as in the proof of Theorem 3,
Section 6, Chapter 5 of \cite{21}  (see also the proof of Lemma
7.3 of \cite{33}). By (the proof of) Theorem  4 and Corollary 1 of
\cite{8} the wave operators $W_\pm (\tau)$ exist and (\ref{1.22})
hold. For this purpose note that  if  $\varphi \in D (H_0), U_0
(t, 0) \varphi$ is globally Lipschitz continuous. Furthermore, by
the proof of Theorem 3.3 given $\varepsilon > 0$, there is $\delta
> 0$ such that

\beq \|(U (t_2, t_1) - I ) \phi \|_{L^2} \leq \varepsilon, \,
\hbox{ if } |t_2 - t_1| \leq \delta, \label{4.79} \ene and where
$\delta$ depends only on the $L^2$ norm of $\phi$. Then, $(U (t,
0))^{-1} U_0 (t, 0) \varphi$ is uniformly continuous if  $\varphi
\in D (H_0)$. We prove that Range $W_\pm (t) = \Hscr_{ac}( U (t +
2 \pi, t))$ arguing as in the proof of Theorem 1.1 of \cite{41}.

\bull
By the intertwining relations \cite{27},
 \beq
 F P_{ac}(F)= \Wscr_{\pm}\, F_0 \, \Wscr^{\ast}_{\pm},
 \label{4.76b}
 \ene
 and, in particular, the absolutely--continuous spectrum of $F$
 coincides with the spectrum of $F_0$, and it is equal to $\ERE$.

As in the proof of Theorem 6.3 of \cite{20} we prove that (see (\ref{1.16})) for $\lambda \in \ERE \backslash \ZETA \backslash \sigma_p (F)$

\beq \hat\Scr (\lambda) = I - 2 \pi i D (\lambda) A G_+ (\lambda)
B D ^\ast (\lambda), \label{4.80} \ene

\beq \hat\Scr (\lambda)^{-1} = I + 2\pi i D (\lambda) A G_-
(\lambda) B D^\ast (\lambda). \label{4.81} \ene $ \hat \Scr
(\lambda)$ is unitary on $\hat\Hscr (\lambda)$. Note that by
(\ref{2.33}) and Lemma 4.1 $D(\lambda) A$ and $D(\lambda) B$ are
compact operators, and hence, $\hat\Scr (\lambda)- I$ is compact.
Suppose that in (\ref{4.1}) $s: = \frac{1 + \varepsilon}{2}
> \frac{3}{2}$ and define,

\beq \psi_m  := V_2 \,\phi_m \label{4.82} \ene with $\phi_m $ as
in (\ref{2.10c}). Since $\phi_m \in \Kscr_{1, - s}, \psi_m \in
\Hscr$. Let us denote

\beq \psi_{-, m} (t, x, \lambda, \nu) = \psi_m (t, x, \lambda,
\nu) - V_2 R_+ (\lambda) V_1 \, \psi_m (\cdot,\cdot, \lambda,
\nu), \label{4.83} \ene
$m < \lambda, \lambda \in \ERE \backslash
\ZETA \backslash \sigma_p (F), \, \nu \in S^2_1$. By taking
adjoint in (\ref{4.33}) and by (\ref{4.38}) and (\ref{4.83}) we
prove that $\psi_{-, m}$ is a solution of the following
Lippmann-Schwinger equation

\beq \psi_{-,m} = \psi_m - V_2 R_{0, +} (\lambda) V_1 \psi_{-,m}.
\label{4.86} \ene Recall that by Lemma 4.1 $V_2 R_{0, +} (\lambda)
V_1 $ is compact in $\Hscr$, and then (\ref{4.86}) has an unique
solution unless there is a non-trivial solution to the homogeneous
equation

\beq \varphi = - V_2 R_{0, +} (\lambda) V_1 \varphi. \label{4.86b}
\ene
But as $V_1 \varphi  = \chi_1 V_1 \varphi $ (\ref{4.86b}) is
equivalent to

\beq \chi_1 \varphi = - Q_{0, +} (\lambda) \chi_1 \varphi,
\label{4.89} \ene and as by (\ref{4.86b}) $\chi_1 \varphi
\not\equiv 0$ , by Lemma 4.2 equation (\ref{4.86}) has an unique
solution for $\lambda \in\ERE \backslash \ZETA$ if and only if
$\lambda \not\in \sigma _p (F)$. Define,

\beq \phi_{-,m}(t,x,\lambda,\nu) = \phi_m(t,x,\lambda,\nu) - R_{0,
+} (\lambda) V_1 \psi_{-,m} (\cdot, \cdot, \lambda, \nu).
\label{4.87} \ene By (\ref{4.87}), and as $\psi_{-,m} = V_2
\phi_{-,m}$,

\beq (F_0 + V) \phi_{-,m}= \lambda \phi_{-,m}.
 \label{4.88} \ene
Observe that $(F_0-\lambda) \phi_{-, m}$ and $V\phi_{-, m}$ belong
to $\Kscr_{- 1, s}$ and that (\ref{4.88}) holds on $\Kscr_{-1,
-s}$.

 Let $T(\lambda)
:= \hat\Scr (\lambda) - I$ be the scattering amplitude. Then, by
(\ref{2.10c}), (\ref{2.17}), (\ref{2.18}), (\ref{4.37}) and
(\ref{4.80}), for $\varphi \in \hat \Hscr (\lambda)$

\beq (T (\lambda) \varphi)_n (\nu) = - 2 \pi i \sum_{m < \lambda}
\int_{S^2_1}\,T_{n,m} (\lambda, \nu, \nu^\prime) \varphi_m
(\nu^\prime) \, d \nu^\prime, \label{4.84} \ene where

\beq T_{n,m } (\lambda, \nu, \nu^\prime) := \left( V \phi_{-,m}
(\cdot, \cdot , \lambda, \nu^\prime) , \phi_n (\cdot, \cdot,
\lambda, \nu)\right). \label{4.85} \ene Our results on the
existence and completeness of the wave operators extend those of
\cite{9}, \cite{41}, to the critical singularity $L^{3/2}$ of the
potential. For other results on scattering with time-periodic
potentials see for example \cite{2}, \cite{19}, \cite{23},
\cite{29}, \cite{40} and \cite{42}.

\section{The Inversion} \sss

We first prepare some results on the Faddeev's Green operator that we need.
For the proofs see \cite{34}.
\begin{enumerate}
\item[1)]
 For $p = p_{\bot} + z \nu \in \CE^3, p_\bot \in \ERE^3, z \in \CE_{\pm}, \nu \in S^2_1,
p_\bot \cdot \nu = 0$, denote,

\beq g_\nu (p) \varphi := \frac{1}{(2 \pi)^{3/2}} \int e^{i k\cdot
x} \frac{1}{(k^2 + 2 p \cdot k)}
 \hat\varphi (k) d k
\label{5.1}
\ene

\end{enumerate}

\noindent where $\hat \varphi = \Fscr_T \varphi$. Then, as a
function of $p_\bot, \nu, z \in \CE_{\pm},\, g_\nu (p)$ is
continuous and it has continuous extensions to $z \in \overline{\CE}_{\pm}$ with values in
$\Bscr (L^2_s, W _{2, - s}), s > 1 /2 $,
with the exception of $(p_\bot, z) = (0, 0)$, and if $s > 1$ also
at $(p_\bot, z) = (0, 0)$.

\begin{enumerate}

\item [2)] For fixed $p_\bot, \nu, g_\nu (p_\bot, z)$ is an
analytic function of $z \in \CE_{\pm}$ with values in $\Bscr (L^2_s,
W_{2, - s})$.

\item[3)] For any $\delta > 0$ there is a constant $C_\delta$ such
that,

\end{enumerate}

\beq \|g_\nu (p) \|_{\Bscr (L^2_s, W_{\rho,- s}}) \leq C_{\delta} (|p_\bot|
+ |z|)^{\rho-1}, 0\leq \rho \leq 2, \label{5.2} \ene for $|p_\bot|
+ | z| \geq \delta, \,s > 1 /2$.

\begin{enumerate}

\item[4)] For $ p_{\nu} \in \ERE$,

\end{enumerate}

$$ g_{\nu, \pm}(p_\bot, p_\nu):= \hbox{s}-\lim_{\epsilon\downarrow
0}g_{\nu}(p_\bot+ (p_{\nu}\pm i\epsilon )\nu )=
$$
$$  e^{-i (p_\bot+p_{\nu}\nu) \cdot x} \left( R_{0,+}((p_\bot+p_\nu
\nu)^2 )- \frac{i}{ 8 \pi^2  |p_\bot + p_\nu \nu|} T^{\ast}(
|p_\bot + p_\nu \nu|)\chi_{\{\pm \omega_{ \nu} > \pm p_\nu
/|p_\bot + p_\nu \nu|\}}\right.
$$
\beq \left.
 T( |p_\bot + p_\nu \nu|)\right) e^{i(p_\bot+p_{\nu}\nu)\cdot
x}.
 \label{5.2b}\ene

 For (\ref{5.2}) with $p^2= 0, p
\not= 0$, see \cite{31}.

Consider now the operator,

\beq g_\nu (p, \gamma)  \varphi:= \frac{1}{(2 \pi)^{3/2}} \int
\frac{e^{i k \cdot x}}{k^2 + 2 p \cdot k  -\gamma} \hat \varphi
(k) d k , \label{5.3} \ene with $p= p_\bot + z \nu, z = \alpha + i
\beta \in \CE_{\pm}$, $\gamma \in \ERE$. If $\gamma \geq 0$,

\beq g_\nu (p, \gamma)  =  e^{i \sqrt{\gamma}\,\omega \cdot x}
g_\nu (p_\bot+ z\nu+\sqrt{\gamma}\omega)   e^{-i
\sqrt{\gamma}\,\omega \cdot x}, \label{5.4} \ene
 where $\omega\in S^2_1$ satisfies,  $\omega\cdot p_\bot = \omega \cdot \nu = 0$. Then, by (\ref{5.2})
for any $\delta > 0$, \beq
\left\|g_{\nu}(p,\gamma)\right\|_{\Bscr\left( L^2_s, W_{\rho,
-s}\right)}\leq
C_{\delta}\left(p^2_{\bot}+\alpha^2+\beta^2+\gamma\right)^{(\rho-1)/2},
0 \leq \rho \leq 2, s > 1 / 2, \label{5.5} \ene for
$\left(p^2_{\bot}+\alpha^2+\beta^2+\gamma\right)^{1/2}\geq
\delta$. In the case where $\gamma < 0$, we prove as in the proof
of Theorem 1.1 and Remark 2.2 of \cite{34} that for any $\delta >
0$ there is a constant $C_{\delta}$ such that,

\beq \left\|g_{\nu}(p,\gamma)\right\|_{\Bscr\left( L^2_s, W_{\rho,
-s}\right)}\leq
C_{\delta}\frac{(1+\left(p^2_{\bot}+\alpha^2+\beta^2+|\gamma|\right)
(|p^2_{\bot}+\alpha^2+\gamma|+\beta^2)^{-1/2}
)^{\rho/2}}{(|p^2_{\bot}+\alpha^2+\gamma|+\beta^2)^{(1-\rho/2)/2}},
0 \leq \rho \leq 2,  \label{5.6} \ene
  $s > 1 / 2$, for $\left(|p^2_{\bot}+\alpha^2+\gamma|+\beta^2 \right)^{1/2}\geq \delta$.

We now proceed as in \cite{32}. For $\delta \in \ERE$ denote,

\beq
\Escr_\delta := \{ \varphi: e^{\delta|x|} \varphi(x) \in L^2\},
\label{5.7}
\ene
with norm

\beq \|\varphi\|_{\Escr_\delta} : = \| e^{\delta|x|}\varphi (x)
\|_{L^2}, \label{5.8} \ene and

\beq \Escr_{1, \delta}: = \{ \varphi \in \Escr_\delta :
\frac{\partial}{\partial x_j} \varphi \in \Escr_\delta, j = 1, 2,
3 \}, \label{5.8b} \ene
with norm

\beq \|\varphi\|_{\Escr_{1, \delta}}: = \left[ \| \varphi
\|^2_{\Escr_\delta}+ \sum^3_{j= 1}\left\| \frac{\partial}{\partial
x_j} \varphi \right\|^2_{\Escr_{\delta}}\right]^{1/2}.
\label{5.9}
\ene

Suppose that $\delta > 0$ and fix $\nu \in S^2_1$.  Then, if
$\gamma > 0$ there is a neighborhood, $O$, of $\ERE^2 \times
\CE\setminus \ERE $, in $\CE^2 \times \CE \setminus \ERE,$ and if
$\gamma \leq 0$ there is a neighborhood, $O$, of $(\ERE^2
\setminus S^2_{\sqrt{|\gamma|}}) \times \CE \setminus \ERE$, in
$\CE^2 \times \CE \setminus \ERE$, such that for any $(p_\bot, z)
\in O$ there is an operator $h_\nu (p_\bot, z, \gamma) \in \Bscr
(\Escr_\delta, \Escr_{ - \delta})$ that is analytic in $O$ and
such that

\beq h_\nu (p_\bot, z, \gamma) = g_\nu (p_\bot + z \nu, \gamma),
p_\bot \in \ERE^2, z \in \CE \setminus \ERE.
\label{5.9b} \ene
Moreover, for fixed $p_\bot, z, \nu$ the family of operators
$h_\nu (p_\bot, z, \gamma)$ is uniformly bounded for $\gamma$ such
that, $|p_{\bot,R}^2+ \gamma| \geq \eta$, for any $\eta > 0$,
where $p_{\bot, R}$ denotes the real part of $p_\bot$.

Designate by $-\Delta_\bot$ the Laplacian on the plane orthogonal to $\nu$ and,

\beq r_{\bot, \pm} (z) := (- \Delta_\bot - z)^{-1}, z \in
\overline{\CE_\pm}\backslash \{0\}. \label{5.10} \ene
The
$r_{\bot, \pm} (z)$ are  integral operators with integral kernel
$\frac{i}{4} H^{(1)}_0 (\sqrt{z} |x-y|)$, with $H^{(1)}_0$ the
modified Hankel function. As is well known, this implies that
$r_{\bot, \pm}(z)$ have analytic continuations across  $(0,
\infty)$ to $|\Im \sqrt{z}|< \delta $ as  operators on $\Bscr
(\Escr_\delta, \Escr_{- \delta})$ for any $\delta >0$, with bound
uniform for $|z| \geq \eta_1, |\Im z| \leq \delta - \eta_2, $ for
any $\eta_1, \eta_2 >0$. Let $f \in C^\infty_0 (\ERE)$ satisfy,
$f(\xi) =1, |\xi| \leq \varepsilon, f (\xi) = 0, |\xi|\geq 2
\varepsilon$ with $\varepsilon$  small enough. Then, $h_\nu
(p_\bot, z, \gamma)$ is defined as

\beq h_\nu (p_\bot, z, \gamma):= h^{(1)}_\nu ( p_\bot, z, \gamma)
+ h^{(2)}_\nu (p_\bot, z, \gamma), \label{5.10b} \ene where,

\beq h_\nu^{(1)} (p_\bot, z, \gamma) := e^{-i p_\bot\cdot x}
\Fscr^{-1}_T \left(\frac{(1-f (k_\nu))}{k^2 +  2 z k_\nu - \gamma
- p^2_\bot} \right) \Fscr_T\, e^{i p_\bot \cdot x},
 \label{5.10c}
\ene
and,

\begin{eqnarray}
& h^{(2)}_\nu (p_\bot, z, \gamma) := e^{-i p_\bot \cdot x
}\Fscr_\nu^{-1} [ r_{\bot, +} ( -k^2_\nu - 2 z k_\nu + \gamma +
p^2_\bot ) \chi_{(-\infty, 0)} (k_\nu) \nonumber  \\ \nonumber & +
r_{\bot, -} ( -k^2_\nu - 2 z k_\nu + \gamma + p^2_\bot) \chi_{(0,
\infty)} (k_{\nu})]
 f(k_\nu) \Fscr_\nu \, \, e^{i p_\bot\cdot x},\\
\label{5.10d}
\end{eqnarray}
where $\Fscr_\nu$ is the Fourier transform along the $\nu$ direction in $\ERE^3$ and
$k_\nu = k \cdot \nu$.

In \cite{32} the case $\gamma = 0$ was considered.

Recall that  $f_1,f_2  \in L^3$ are compact operators  from
$W_{1,2}$ into $L^2$. Then,
$Q_{\nu}(p_\bot,z,\gamma):=e^{-\delta_0 |x|}\,f_1\,
h^{(1)}_{\nu}\break e^{-\delta_0 |x|}f_2$ is compact in $L^2$, and
its norm tends to zero as $ \gamma \to - \infty$. In the case $
\gamma
> 0$ we write, $z\nu +p_\bot= q + z_1 \mu$, with $q \in \ERE^3,
z_1 \in \CE \setminus \ERE, \mu \in S^2_1, q \cdot \mu=0$ and $\Im
z_1 > 0$ . Then, \beq
 Q_{\nu}(p_\bot,z,\gamma)= f_1(x)\, e^{i\sqrt{\gamma}\omega\cdot x}e^{-\delta_0 |x|}
g_{\mu}(p) \Fscr_{\nu}^{-1}\, (1- f(k_\nu))\Fscr_{\nu}
 e^{-\delta_0 |x|}\,e^{-i\sqrt{\gamma}\omega\cdot x} f_2(x),
\label{5.10e} \ene where, $p:= q + z_1 \mu+ \sqrt{\gamma}\omega$
with $ \omega \in S^2_1, q \cdot \omega =\mu\cdot\omega= 0$.
$\Fscr_{\nu}^{-1}\, (1- f(k_\nu))\Fscr_{\nu}$ is bounded in
$L^2_s$. Moreover, for $\phi, \psi$ in Schwartz space, $\left(
Q_{\nu}(p_\bot,z,\gamma) \phi,\psi\right)$ is analytic in $z_1$
and it tends to zero as $\Im z_1 \to \infty$. Hence, it follows
from the maximum-modulus principle ( see \cite{47}, page 231) that
it takes the maximum for $z_1$ real.

Moreover (see the proof of Lemma 4.1), $f_1 r_{0,\pm}(z) f_1$ is
bounded in $L^2$ with bound uniform for $ z \in
\overline{\CE_\pm}$. Denote, $d(\rho):= (1/ \sqrt{2} \rho^{1/4})
\, T(\sqrt{\rho})$ with $T(\rho)$ the trace operator (\ref{2.13}).
As,

\beq f_1\,  d^{\ast}(\rho)\, d(\rho) f_1 := \frac{1}{2\pi
i}f_1\left[ r_{0,+}(\rho)- r_{0,-}(\rho) \right] f_1,
\label{5.10g} \ene we have that $d(\rho)f_1 \in \Bscr(L^2, L^2
(S^2_1))$ with bound uniform in $\rho$. Moreover, if $ f_1\in
C^{\infty}_0, f_1 \,r_{0,\pm}(\rho)\, f_1$ is Hilbert-Schmidt, and
hence, in this case, $ d(\rho)f_1$ is compact. Approximating $f_1$
by functions in $C^{\infty}_0$ we prove that $d(\rho)f_1$ is
compact for $f_1 \in L^3$.

Furthermore, for $z_1$ real, $\Im z= \Im p_\bot =0$ and it follows
from (\ref{5.2b}) that in this case,

\beq \left|\left( Q_{\nu}(p_\bot,z, \gamma )\phi, \psi
\right)\right|\leq C \|f_1\|_{L^3}\, \|f_2\|_{L^3}\,
\|\phi\|_{L^2} \|\psi\|_{L^2}, \label{5.10h} \ene uniformly in
$\gamma > 0$. Here we use that $\Fscr_\nu^{-1}
(1-f(k_{\nu}))\Fscr_{\nu}$ is bounded in all $L^p$ spaces. By the
maximum-modulus argument above, this is also true for all $z \in
\CE$, and by
 continuity, it also holds for all $ \phi, \psi \in L^2$, and it
follows that $Q_{\nu}(p_\bot,z,\gamma)$ is uniformly bounded in
$L^2$ for all $\gamma >0$. Note that if $f_1,f_2 \in
C^{\infty}_0$, the norm of $Q_{\nu}(p_\bot,z, \gamma )$ goes to
zero as $ \gamma \to \infty$. Hence, approximating $f_1,f_2$ by
functions in $C^{\infty}_0$ we prove that this is also true for
$f_1,f_2 \in L^3$.

The function $H_0^{(1)} (z)$ satisfies the following estimate
\cite{6}, \beq |H^{(1)}_0 (z) | \leq C\cases{ |\ln z|, &  $|z|
\leq  1 / 2, $ \cr \frac{e^{|\Im z|}}{|z|^{1/2}}, &  $|z|  \geq  1
/ 2. $\cr } \label{5.10i} \ene From this it follows that
$e^{-\delta_0 |x|}\,f_1\, h^{(2)}_{\nu}\, e^{-\delta_0 |x|}f_2$ is
Hilbert-Schmidt in $L^2$  and that its norm goes to zero as $
|\gamma| \to \infty$.

 For any  $\delta\in \ERE$, we define, \beq \Dscr_\delta : =  L^2
(T, \Escr_\delta). \label{5.11} \ene For \beq
 (p_\bot, z) \in O_{\ZETA}:=\left\{(p_{\bot},z)\in O, \, \hbox{with}\,
p_{\bot, R}^2 \not \in \ZETA\right\}, \label{5.11c} \ene
let us
define,

\beq H_\nu (p_\bot, z, m) := \oplus_{n=-\infty}^{\infty} h_\nu
(p_\bot, z, m - n)\, P_n \in \Bscr (\Dscr_{\delta},
\Dscr_{-\delta}). \label{5.12} \ene
If
 $V$ satisfies the conditions of  Theorem 1.1, we prove as in Lemma 4.1- using the results above- that

\beq M_\nu (p_\bot, z, m) := V_2 H_\nu (p_\bot, z, m) V_1,
(p_\bot, z) \in O_{\ZETA}, \label{5.15} \ene are compact operators
on $\Hscr$ and, moreover, that $M_\nu (p_\bot, z, m)$ go to zero
in norm as  $p_{\bot}\in \ERE^2, |\Im z | \to \infty$, uniformly
in $p_{\bot}$. Then, $(I+ M_\nu (p_\bot, z, m))$ has a bounded
inverse in $\Hscr$ for $(p_\bot, z) \in O \backslash O_e (m)$,
where the exceptional set $O_e (m)$ has the following properties.
The intersection of $O_e(m)$ with $\{(p_{\bot},z)\in O_{\ZETA}:
p_{\bot}\in \ERE^3, |\Im z| \geq M \}$ is empty for some $M >0$. Let
$U \subset \CE_{\pm}$ be any open set such that there is a sequence,
$z_n \in U$ with $|\Im z_n| \to \infty$ as $ n \to \infty$.
Moreover, let $ z \in U, \rightarrow p_{\bot}(z)$ be any analytic
function such that $(p_{\bot}(z),z)\in O_{\ZETA}, z \in U$ and
that $p_{\bot}(z_n)\in \ERE^3$. Then, the set $\{z\in U :
(p_{\bot}(z),z ) \in O_e(m)\}$ has no accumulation points in $U$.

Let us now define for $\lambda \in \ERE \backslash \ZETA \backslash \sigma_p (F),
\psi_{+, m} $ as in (\ref{4.83}), but with $R_-(\lambda)$ instead of $R_+ (\lambda)$,

\beq \psi_{+, m} (t, x, \lambda, \nu) := \psi_m (t, x, \lambda,
\nu) - V_2 R_- (\lambda) V_1 \psi_m (\cdot, \cdot, \lambda, \nu),
\label{5.16} \ene and (see (\ref{4.87}))

\beq \phi_{+, m} (t, x, \lambda, \nu) := \phi_m (t, x, \lambda,
\nu) - R_{0, -} (\lambda) V_1 \psi_{+, m} (\cdot, \cdot, \lambda,
\nu). \label{5.17} \ene
 We define for $g \in \hat{\Hscr}
(\lambda)$,
\beq \phi_{\pm, g} (t, x, \lambda) := \sum_{m <
\lambda} \int_{S^2_1}\phi_{\pm, m} (t, x, \lambda, \nu) g_m (\nu)
d\nu. \label{5.18} \ene

Suppose that we are given another potential, $\tilde V$, that satisfies the
conditions of Theorem 1.1, and let us denote by $\psi^{\sim}_{\pm, m}, \phi^{\sim}_{\pm, m},
 S^\sim (\lambda), \tilde T_{n, m}(\lambda)$ the corresponding quantities for
 $\lambda \in \ERE \backslash \ZETA \backslash \sigma_ p (\tilde F)$.
It follows from (\ref {4.83}), (\ref {4.87}), (\ref {5.16}), and
(\ref{5.17})  that,

\begin{eqnarray}
2 \pi i \bigg[
( ( F_0 - \lambda) \phi_{-,n} , \tilde\phi_{+, m}) &-&  (\phi_{-, n} , (F_0 - \lambda)
\tilde \phi_{+, m})
\bigg] \nonumber \\
= - 2 \pi i \bigg[ ( V \phi_{-, n}, \phi_m) &-&  ( \tilde V
\tilde\phi_{-, n},
 \phi_m)\bigg] . \\
\label{5.19}
\end{eqnarray}
Then, by  (\ref {4.84}) and (\ref{4.85})

\beq 2\pi i \bigg[ ((F_0 - \lambda) \phi_{-, g} , \tilde \phi_{+,
\tilde g}) - (\phi_{-,g}, (F_0 - \lambda) \tilde\phi_{+,\tilde g}
)\bigg] = \left((S (\lambda) - \tilde S (\lambda)) g, \tilde
g\right), g , \tilde g \in \hat\Hscr (\lambda). \label{5.20} \ene

Moreover, for $(p_\bot, z) \in O \backslash  O_e (m)$ let us
denote,
\beq \Omega_{m, \nu} (t, x, p_\bot, z) := e^{i (p_\bot + z
\nu)\cdot x} [e^{imt}- H_\nu (p_\bot, z, m) V_1 \Gamma_{m,\nu}
(\cdot, \cdot , p_\bot, z)], \label{5.21} \ene where,

\beq
\Gamma_{m,\nu} (t, x, p_\bot, z) := (1+ V_2 H_\nu (p_\bot, z, m) V_1)^{-1}e^{i m t} V_2,
\label{5.22}
\ene
is the unique solution of the Lippmann-Schwinger equation,

\beq \Gamma_ {m,\nu} (t, x,  p_\bot, z) = e^{imt} V_2 - V_2 H_\nu
(p_\bot, z, m) V_1 \Gamma_{m,\nu} (\cdot, \cdot, p_\bot, z)
\label{5.23} \ene in $\Hscr$.
 As $e^{-\delta |x|}
H_{\nu}(p_{\bot}, z,m) V_1 \in \Bscr(\Hscr)$ for $\delta > 0$ we
have that, \beq e^{-\delta |x|} H_{\nu}(p_{\bot}, z,m)\,
V_1\,\Gamma_{m,\nu}(\cdot,\cdot,p_{\bot},z)\in \Hscr.
\label{5.23b} \ene Then, $\Omega_{m,\nu} \in \Dscr_{-\delta}$, if
$\delta >  |\Im z|+ |\Im p_{\bot}|$. Moreover,  if $(p_{\bot}+ z
\nu)^2= \lambda-m$, $(F_0-\lambda)\,\Omega_{m,\nu} \in \Dscr_{-1,
\delta_0}, V_2\, \Omega_{m,\nu}\in \Dscr_{\delta_0}$, and then, $V
\, \Omega_{m,\nu} = V_1 V_2\, \Omega_{m,\nu}\in \Dscr_{-1,
2\delta_0}$.  Hence, \beq (F_0 +V-\lambda) \Omega_{m,\nu}=0,
\label{5.24} \ene in $ \Dscr_{-1,-\delta}$.

By (\ref{4.35}) and using (\ref{4.38}), (\ref{4.83}),
 (\ref{4.87}),
 (\ref{5.16}) and
 (\ref{5.17}), we see that

\beq \phi_{\pm, m}(t,x,\lambda,\nu) = \phi_m(t,x,\lambda,\nu) -
R_{\mp} (\lambda) V_1 \psi_m(\cdot, \cdot , \lambda, \nu).
\label{5.25} \ene Let $Q$ be any function  that  satisfies the
conditions for $V_j,j=1,2$ in Theorem 1.1. Let $f \in \Hscr$, and
suppose that

\beq
(f, Q \phi_{+, g}) = 0,
\label{5.26}
\ene
for all $g \in \hat\Hscr(\lambda)$. Denote,

\beq h:=(I-A G_- (\lambda) B R_{0,_-} (\lambda) )Q f. \label{5.27}
\ene Then (see (\ref{2.10c}) and (\ref{2.17})), for $\lambda > m,
g_m \in L^2 (S^2_1)$,

\beq
(T_m (\lambda) h, g_m) = \int_{S^2_1} \bar g_m (\nu) (f, Q\, \phi_{+, m}
(\cdot, \cdot, \lambda, \nu)) d \nu =0,
\label{5.28}
\ene
where we have taken adjoints in (\ref{4.33}) and we used (\ref{4.37}), and
(\ref{5.25}). It follows that if
(\ref{5.26}) holds then (see (\ref{2.18})),

\beq
D (\lambda) h = 0.
\label{5.29}
\ene

Let us designate,

\beq u_-:= R_-(\lambda) Qf. \label{5.30} \ene By (\ref{4.34}) and
(\ref{4.38}),

\beq
u_- = R_{0, _-} (\lambda) h \in \Hscr_{-s}, s > 1/2.
\label{5.31}
\ene
Then, by (\ref{5.29})  it follows as in the proof of Lemma 4.2 that $u_- \in \Hscr$.
Note that
$h \in \Escr_{-1,\delta}, \delta < \delta_0$.
By (\ref{5.31}), as $u_- \in \Hscr$ and
(\ref{5.29}) holds,

\beq (\tilde \Fscr u_-)_m (k) = \frac{1}{k^2 - \lambda + m}
(\tilde\Fscr h)_m (k), \label{5.32} \ene with $(\tilde\Fscr h)_m
(\sqrt{\lambda-m}\,\nu) = 0, \nu \in S^2_1, \lambda > m$. It
follows from the Paley--Wiener theorem  \cite{25} that $u_- \in
\Escr_{1, \delta}$, for any $ \delta < \delta_0$.

Arguing as in the beginning of the proof of Lemma 4.2 we obtain
that,

\beq
(F_0 + V-\lambda) u_- = Q f.
\label{5.33}
\ene

For $\delta < \delta_0$ denote,

\beq N := \left\{ \varphi \in \Hscr :\varphi = Q \phi, \phi \in
\Escr_{- \delta}, (F_0 - \lambda ) \phi \in \Dscr_{-1, - \delta},
V \phi \in \Dscr_{-1, -  \delta,} \hbox{ and } (F_0 + V- \lambda)
\phi = 0 \right\}. \label{5.34} \ene
Then, for any $\varphi \in
N:$

\beq (f, \varphi) = ((F_0 + V- \lambda)u_ -, \phi) = 0.
\label{5.35} \ene
This implies that $f \in N^\bot$, and hence, the
set \beq \left\{\varphi: \varphi= Q\phi_{+, g}, \,\hbox{with}\, g
\in \hat\Hscr (\lambda)\right\} \label{5.35b} \ene
 is dense in the closure, $\overline{ N}$, of $N$ in $\Hscr$.
We prove that the set
\beq
\left\{\varphi: \varphi= Q\phi_{-, g},
\,\hbox{with}\, g \in \hat \Hscr (\lambda)\right\}
\label{5.35c}
\ene
is dense in $\overline{ N}$ in the same way.

We argue now as in  \cite{32}. For any $m_1, m_2 \in \ZETA$ denote $m_0 = \max [m_1, m_2]$.
Suppose that $\lambda > m_0$.
 Then, for any $k \in \ERE^3$ take $\nu, \omega \in S^2_1$ with $k\cdot\nu = k\cdot \omega = \nu\cdot \omega =0$,
and for $\rho >0$ such that $\rho^2+\lambda -m_j \notin \ZETA, j=1,2$, define

\beq p_\bot  := \frac{k}{2}+(\rho^2- \frac{1}{4} k^2 + \lambda -
m_1)^{1/2} \omega, \label{5.37} \ene

\beq p^\prime_\bot  := -\frac{k}{2} + (\rho^2 - \frac{1}{4}k^2 +
\lambda- m_2)^{1/2} \omega, \label{5.38} \ene for $\rho > \left[
\max( \frac{1}{4} k^2 - \lambda + m_0, 0)\right]^{1/2}$. The
integral,

\beq I:=\int (V- \tilde V) (t, x) \Omega_{m_1,\nu} (t, x, p_\bot,
i\rho) \overline{\tilde{\Omega}_{m_2,\nu} (t, x,
{p_\bot^{\prime}}, -i\rho)}\, dt \ dx \label{5.36} \ene converges
and it is meromorphic for $\rho$ in a neighborhood of
 $\rho > \left[ \max( \frac{1}{4} k^2 - \lambda + m_0, 0)\right]^{1/2}$.
 Moreover, by (\ref{5.20}) and the density argument above, $I=0$ for $ \rho < \delta_0$.
If for some  $\epsilon $ with, $ \delta_0 > \epsilon > 0$, $k^2 <
4 \left[(\delta_0 - \epsilon)^2 + \lambda - m_0\right]$
 this contains all the $\rho $ as above with $\delta_0- \epsilon < \rho < \delta_0$.
It follows by analyticity in $\rho$ that for these $k $, $I=0$
for $\rho
> \left[ \max( \frac{1}{4} k^2 - \lambda + m_0, 0)\right]^{1/2}$.
Taking the limit $\rho  \to \infty$ we have that,

\beq
\int e^{i k \cdot x} e^{i (m_1 - m_2) t}[V (t, x) - \tilde V (t, x)] dt \ dx = 0.
\label{5.39}
\ene
But as the Fourier transform of $V-\tilde V$ is analytic in $k$ for $|\Im k| < 2 \delta_0$,
this holds for all
$k \in \ERE^3, m_1, m_2 < \lambda$.
This implies that $V-\tilde V \equiv 0$, and it completes the proof of Theorem 1.1.

\end{document}